\documentclass[lettersize,journal]{IEEEtran}
\usepackage{cite}
\usepackage{amsmath,amssymb,amsfonts}
\usepackage{bm}
\usepackage{mathtools}
\usepackage{algorithmic}
\usepackage{graphicx}
\usepackage{textcomp}
\usepackage{xcolor}
\usepackage{booktabs}
\usepackage{multirow}
\usepackage{multicol}
\usepackage{xspace}
\usepackage{xurl}
\usepackage{tikz}
\usepackage{environ}

\usepackage{subfig}
\usepackage{rotating}
\usepackage{amssymb}
\usepackage{pifont}
\newcommand{\cmark}{\ding{51}}%
\newcommand{\xmark}{\ding{55}}%
\newcommand*\emptycirc[1][1ex]{\tikz\draw (0,0) circle (#1);} 
\newcommand*\halfcirc[1][1ex]{%
  \begin{tikzpicture}
  \draw[fill] (0,0)-- (90:#1) arc (90:270:#1) -- cycle ;
  \draw (0,0) circle (#1);
  \end{tikzpicture}}
\newcommand*\fullcirc[1][1ex]{\tikz\fill (0,0) circle (#1);}

\def\BibTeX{{\rm B\kern-.05em{\sc i\kern-.025em b}\kern-.08em
    T\kern-.1667em\lower.7ex\hbox{E}\kern-.125emX}}

\newif\ifdraft
\draftfalse

\ifdraft
\newcommand{\nojan}[1]{\textcolor{red}{{\sf (NS:} {\sl{#1})}}}
\newcommand{\anees}[1]{\textcolor{blue}{{\sf (AA:} {\sl{#1})}}}
\else
\newcommand{\nojan}[1]{}
\newcommand{\anees}[1]{}
\fi

\newcommand{\Prv}{$\mathcal{P}$\xspace}
\newcommand{\Vrf}{$\mathcal{V}$\xspace}
\newcommand{\Cir}{$\mathcal{C}$\xspace}

\begin{document}

\title{Zero-Knowledge Proof Frameworks: A Systematic Survey}
\author{\textnormal{Nojan Sheybani$^1$, Anees Ahmed$^2$, Michel Kinsy$^2$, Farinaz Koushanfar$^1$} \\
$^1$UC San Diego, $^2$Arizona State University \\
$^1$\{nsheyban, farinaz\}@ucsd.edu, $^2$\{aahmed90, mkinsy\}@asu.edu}



\maketitle

\begin{abstract}
Zero-Knowledge Proofs (ZKPs) are a cryptographic primitive that allows a prover to demonstrate knowledge of a secret value to a verifier without revealing anything about the secret itself. ZKPs have shown to be an extremely powerful tool, as evidenced in both industry and academic settings. In recent years, the utilization of user data in practical applications has necessitated the rapid development of privacy-preserving techniques, including ZKPs. This has led to the creation of several robust open-source ZKP frameworks. However, there remains a significant gap in understanding the capabilities and real-world applications of these frameworks. Furthermore, identifying the most suitable frameworks for the developers' specific applications and settings is a challenge, given the variety of options available. The primary goal of our work is to lower the barrier to entry for understanding and building applications with open-source ZKP frameworks.

In this work, we survey and evaluate 25 general-purpose, prominent ZKP frameworks. Recognizing that ZKPs have various constructions and underlying arithmetic schemes, our survey aims to provide a comprehensive overview of the ZKP landscape. These systems are assessed based on their usability and performance in SHA-256 and matrix multiplication experiments. Acknowledging that setting up a functional development environment can be challenging for these frameworks, we offer a fully open-source collection of Docker containers. These containers include a working development environment and are accompanied by documented code from our experiments. We conclude our work with a thorough analysis of the practical applications of ZKPs, recommendations for ZKP settings in different application scenarios, and a discussion on the future development of ZKP frameworks.
\end{abstract}


\section{Introduction}

Privacy-preserving cryptographic methods have become increasingly vital as privacy and data security evolve into a higher priority in new applications. Zero-Knowledge Proofs (ZKPs) enable a prover \Prv to prove to a verifier \Vrf that a statement is true, without revealing any information beyond the validity of the statement itself. While ZKPs are most prominently known to the general public in blockchain applications \cite{boo2021litezkp, chainOverviewZeroKnowledge, binanceWhatZeroknowledge, vcapko2022state}, they have also been effectively applied in many other real-world domains, such as healthcare \cite{tomaz2020preserving, sharma2020blockchain, gaba2022zero}, traditional finance \cite{rabin2012strictly, thorpe2009zero, lin2021efficient}, and government \cite{bamberger2022verification, landau1988zero}. ZKPs are an excellent solution for verifying data and computation in a secure fashion, however there are still many challenges before they can become a practical privacy-preserving solution.

Although introduced in the 1980s \cite{goldwasser2019knowledge}, recent algorithmic and computing advances have garnered the evolution of ZKPs from a theoretical construct to a relatively practical cryptographic primitive. ZKPs have garnered the interest of researchers and developers as concerns over data privacy grow, which has caused significant improvements in both theory and implementation. The first significant milestone in the practical application of ZKPs was the development of zero-knowledge succinct non-interactive arguments of knowledge (zk-SNARKs) introduced by Ben-Sasson et. al \cite{ben2014succinct} in 2013. 

In the decade since the introduction of zk-SNARKs, the zero-knowledge landscape has evolved to include a diverse set of ZKP constructions, such as zk-STARKs \cite{ben2018scalable}, which build off of zk-SNARKs and are discussed at length in this work. For many of the ZKP constructions that are available, there are several prominent frameworks, stemming from industry and academia, that allow developers to create their own ZKP applications. Despite the availability of these open-source frameworks and the demand of privacy-preservation in real-world systems, the implementation of ZKPs in practical applications has been limited. This can be attributed to three ongoing challenges: 1) performance; 2) usability; 3) accessibility. These are the attributes that we evaluate the chosen open-source frameworks on in this work.
The journey towards making ZKPs a de facto solution for privacy-preserving applications is hamstrung by the performance, due to the complexity of current accessible ZKP protocols. 
In this work, we hope to find the schemes with the best performance for each type of ZKP construction, evaluated over several metrics on CPU and we will discuss the next steps that can be taken towards practical ZKP adoption. 

Usability and accessibility are common problems that face privacy-preserving technologies, especially those stemming from academia, and the ZK landscape is no different. Due to the surge of research that has been done in the ZK space, there has been a sudden increase in the number of available frameworks for developers. For a nascent developer of ZKP applications, especially one with little exposure to cryptography and ZK concepts, this can seem like a near-impossible field to navigate. Even for experienced developers, these frameworks are often hard to use, due to their (mostly) poor documentation or reproducible examples. While this is understandable in academic settings, due to time and resource constraints, a significant step towards enabling practical ZKP usage is demystifying the currently existing frameworks and lowering the barrier of entry for experienced and unexperienced developers. Alongside this, it is also currently difficult for developers to decipher whether a framework is usable for their custom application, due to the different ZKP constructions available, each with different underlying arithmetic, security guarantees, and interaction/communication requirements. 

While there is no arguing that the development of the open-source ZKP frameworks has significantly reduced the amount of necessary effort for building new applications, the field is still difficult to navigate. The available open source frameworks have been used to enable secure verification of computation, data, and identity in the domains of machine learning \cite{zhang2020zero}, networking \cite{grubbs2022zero}, IP protection \cite{sheybani2023zkrownn}, and many more. Although there has been this evident uptick in ZKP frameworks and applications, there is no overview of the ZK landscape that is both cryptographer and non-cryptographer friendly. Alongside this, it is hard to find a clear path for where ZKPs can be improved so they can be more broadly integrated into practical real-world applications. 

This paper aims to provide users with a guide to ZKPs and the available ZKP frameworks, allowing readers to gain a high-level overview of the ZK landscape, while also providing new quantitative benchmarks and details for developers to choose the best ZKP framework for their application. To achieve this goal, we conduct an extensive survey of the ZK landscape, gathering several state-of-the-art frameworks representing the seminal ZKP constructions. We first evaluate these existing tools based on the usability and accessibility of their repositories for a non-experienced cryptographic application developer, highlighting their features and shortcomings from a design standpoint. 
We then evaluate a subset of the most accessible and usable frameworks, primarily those that expose a high-level API, based on their performance through an in-depth analysis of their runtime and communication complexity. Performance is measured over two custom benchmarks that represent commonly used functions in privacy-preserving computation: matrix multiplication and SHA-256 compression. 

We provide a discussion of the different constructions at a high-level to guide developers in their choice of framework. Our experimental evaluation, insights, and recommendations should provide a general guide to developers on how to whittle down the available frameworks to ones that fit their application setting, bandwidth, and computational requirements.
We conclude our work with a discussion on some of the cutting-edge applications that ZKPs have been utilized in, the challenges that ZKP applications currently face, and the future of ZKPs.

Unfortunately, many of the prominent works that provide open-source frameworks do not include a proper documentation or reproducible examples, thus hindering developers in integrating these frameworks into their applications. Alongside this, many of the frameworks require complex local environment build dependencies. To combat this hurdle, we provide a new open-source Github repository\footnote{\url{https://github.com/ACESLabUCSD/ZeroKnowledgeFrameworksSurvey}} containing all the tools necessary to build a custom ZKP application with any framework discussed in this survey. Not only do we provide open-source Docker environments for each frameworks with reproducible documented examples, but we also include Docker containers for other helpful tools, such as circuit building and inspection tools. This repository is also well-documented and actively maintained to encourage users to immediately start building custom applications, rather than focusing on setup troubles.

The goal of this survey is to lower the barrier of entry to building ZKP applications by providing an in-depth overview of ZKPs, the existing constructions, the available open-source frameworks and their capabilities, and the usability, accessibility, and performance of each available framework. This paper is written so that a reader with no prior knowledge of ZKPs can garner a high-level understanding of the landscape, while experienced readers can sharpen their knowledge of the details of ZKPs and gain insights on the available tools for ZK-based application development. In short, our scientific contributions are:
\begin{itemize}
    \item We present the first survey of open-source ZKP frameworks, spanning \textit{all} ZKP constructions, with accompanying open-source environments for each framework, including benchmarks and documentation.
    \item We perform extensive analysis of select open-source ZKP frameworks on scalability, runtime, and proof size on two benchmarks representing prominent domains of ZKPs in current practice.
    \item We provide a thorough analysis of the capabilities, usability, and accessibility of each open-source ZKP framework.  Based on the insights of our work, we customize suggestions of frameworks for different use cases based on available compute power, developer experience, and application type. Finally, we provide novel insights on the current state of ZKP and the necessary path to further boost practicality.
\end{itemize}
\nojan{add scientific contribution to this DONE}

\subsection{Related Work}

To the best of our knowledge, this work is the first to systematically survey and benchmark open-source ZKP frameworks spanning \textit{all} constructions for practical settings and realization. \cite{CelerNetwork2023Pantheon, Delendum2023ZKSystemBenchmarking} have considered a very limited amount of frameworks, but their industry-led work is largely comparing different constructions to each other (zk-SNARK vs. zk-STARK), rather than comparing frameworks of the same construction to each other (zk-SNARK vs. zk-SNARK). While there are very interesting insights made, we believe that our work is much more objective, systematic and extensive, while also adding the element of usability and accessibility analysis.
Another survey on ZKP frameworks has been conducted \cite{9520375}, however this work only focuses on zk-SNARKs and does not look into the usability, accessibility, or performance of the chosen frameworks. Also, the work is largely focused on the application of zk-SNARKs in the blockchain. Similarly, \cite{partala2020non} conducts a survey on non-interactive ZK applications in the blockchain. The main focus of this work is the analysis of privacy-protection schemes for smart contracts. This work approaches ZK scheme analysis from a theoretical standpoint. Rather than focusing on usability and accessibility, this work primarily focuses on analyzing the asymptotics of the available schemes. Unfortunately, this is not fully representative of the performance of these schemes in practice, due to some schemes having high constants in their asymptotic complexities. While we believe that surveying zk-SNARKs is very important, we note that many zk-SNARK schemes are not post-quantum secure. As post-quantum security becomes a rapidly growing concern, our work purposefully inspects every available ZKP construction to provide insights into post-quantum secure frameworks, alongside more established zk-SNARK frameworks. We believe that limiting our work to zk-SNARKs would not be fully representative of the ZK landscape.

We model our paper after the seminal surveys in privacy-preserving technology centered around MPC \cite{hastings2019sok} and FHE \cite{viand2021sok}. 
Like these works, we aim to provide as detailed of a description as we can surrounding the usability, accessibility, and performance of our chosen frameworks, while providing a digestible guide for developers choosing a tool for their ZK-based applications.
\section{Zero-Knowledge Proofs}

Zero-Knowledge Proofs (ZKPs) are a cryptographic primitive that allow a prover \Prv to prove to a verifier \Vrf that they know a secret value $w$, called the witness, without revealing anything about $w$. \Prv does this by showing that they know a secret value $w$ such that $\mathcal{F}$ evaluated at $w$ equals some public output $y$. Formally, \Prv sends a proof attesting that $\mathcal{F}(x; w)=y$, where $x$ and $y$ are public inputs and outputs, respectively. ZKPs have three core attributes \cite{goldreich1994definitions}:
\begin{enumerate}
    \item \textbf{Soundness}: \Vrf will find out, with a very high probability, if a \Prv is dishonest if the statement is false.
    \item \textbf{Completeness}: An honest \Prv can convince \Vrf if the statement is true.
    \item \textbf{Zero-Knowledge}: If the statement is true, \Vrf will learn nothing about the \Prv's private inputs - only that the statement is true.
\end{enumerate}
In the following sections, we discuss the evolution of ZKPs, the nuances of specific classes and schemes, and provide a detailed overview of the current ZK landscape.

\subsection{Taxonomy of ZKPs}

In this work, we analyze 25 ZK protocols. Amongst these protocols are a mix of interactive and non-interactive schemes. An in-depth explanation of the difference between interactive and non-interactive schemes can be found in Appendix \ref{sec:interactive}. From now on, we describe computation as circuits \Cir, as that is what they are referred to as in ZK literature. This is due to the process of arithmetization, which represents functions, such as Python/C++ code, as arithmetic circuits, then converts these circuits into a mathematical representation (e.g. polynomials) that can be used within ZKPs. Oftentimes, an intermediate step between the input and output is a set of constraints that describes the code/circuit. These constraints act as the basis for the mathematical representation. For brevity's sake, we do not discuss the details of arithmetization and refer to the brilliant explanations of \cite{LambdaClass2023ArithmetizationSchemes, ButerinQuadraticArithmeticPrograms}. In this text, we only treat arithmetization as a black-box and do not require the knowledge of specific details, only the inputs (e.g. code) and outputs (e.g. mathematical representation). Table \ref{tab:pros} compares the seminal ZK protocols at a high-level. Below, we describe the taxonomy of the general schemes that underlie our chosen ZK protocols in detail. 

\begin{table*}[t]
\centering\resizebox{\textwidth}{!}{
\begin{tabular}{lll}
\toprule
\textbf{Construction} & \textbf{Key Advantages} & \textbf{Key Disadvantages}\\
\midrule
zk-SNARKs & Succinct, Publicly Verifiable & Trusted Setup Required, Computationally Expensive to Prove, Not Post-Quantum 
\\
\midrule
zk-STARKs & No Trusted Setup, Post-Quantum Secure, Scalable Prover, Publicly Verifiable & Larger Proof Sizes, Slow Verification 
\\
\midrule
MPCitH & No Trusted Setup, Post-Quantum Secure, Publicly Verifiable & Slow Verification, Computationally Expensive Proving
\\
\midrule
VOLE-ZK & Highest Scalability, No Trusted Setup, Post-Quantum Secure & Slow Verification, Designated Verifier
\\
\bottomrule
\end{tabular}}
\caption{Core Attributes of Popular ZKP Constructions}
\label{tab:pros}
\end{table*}

\begin{table}[t]
    \centering\centering\resizebox{\columnwidth}{!}{
\begin{tabular}{l|cccc}
\hline
& zk-SNARKs & zk-STARKs & MPCitH & VOLE-ZK \\
\hline
Prover complexity & $O(n\log{}(n))$ & $O(n\text{poly-log}(n))$& $O(n)$ & $O(n)$ \\
\hline
Verifier complexity & $O(1)$ & $O(\text{poly-log}(n))$ & $O(n)$ & $O(n)$ \\
\hline
Proof size & $O(1)$ & $O(\text{poly-log}(n))$ & $O(n)$ & $O(n)$ \\
\hline
Trusted setup & \cmark & \xmark & \xmark & \xmark \\
\hline
Non-interactive & \cmark & \cmark & \cmark & \xmark \\
\hline
Post-quantum secure & \xmark & \cmark & \cmark & \cmark \\
\hline
Practical proof size & ~120-500 bytes & ~10 KB - 1 MB & ~10-1000 KB & ~5-200 KB \\
\hline
\end{tabular}}
\caption{Asymptotic attributes of presented ZKP constructions. We do note that, due to the variance of schemes within each construction, the algorithmic complexities are generalized and may not hold true for all schemes within a given construction.}
\end{table}

\textbf{Zero-Knowledge Succinct Non-Interactive Arguments of Knowledge (zk-SNARKs)} are, as the name suggests, a class of non-interactive protocols that boast small proof size \cite{ben2014succinct}. Although ZKPs were originally conceived in the late 1980's \cite{goldwasser2019knowledge}, zk-SNARKs were formally introduced about a decade after. Efficient instantiations of zk-SNARKs were introduced in the last decade, resulting in recent advancements in making zk-SNARKs practical and efficient for widespread use. 
This means there are much more mature open-source and real-world implementations available. The most common forms of zk-SNARKs are referred to as \textit{pre-processing zk-SNARKs}. One of the main drawbacks of these zk-SNARKs are that they require a trusted setup for every new circuit \Cir, which is computationally intensive and requires communication of large proving and verifying keys to the respective parties. 
Alongside this, \Prv must normally be computationally powerful in order to ensure small proof size. This is due to the fact that most zk-SNARKs are reliant on elliptic curve cryptography (ECC) as their underlying cryptographic arithmetic.
Recent works have introduced zk-SNARKs that can utilize \textit{universal} trusted setups \cite{plonk, chiesa2020marlin} for established maximum circuit sizes, and zk-SNARKs that do not require a trusted setup at all \cite{setty2020spartan, wahby2018doubly}. 
Due to the non-interactivity, zk-SNARKs are \textit{publicly verifiable}, meaning any verifier can verify them without recomputing the proof. One of the most common underlying schemes, especially in our highlighted frameworks, for zk-SNARKs is Groth16 \cite{groth16}, which improves upon the original Pinocchio \cite{parno2016pinocchio} protocol. zk-SNARK arithmetization \textit{typically} results in a set of constraints, called Rank 1 Constraint Systems (R1CS) \cite{belles2022circom}, which are then converted to a set of polynomials, called a Quadratic Arithmetic Program (QAP) \cite{gennaro2013quadratic}. We do note that there are different formats that are zk-SNARKs are compatible with, such as Algebraic Intermediate Representations (AIR) \cite{ben2018scalable} and Plonkish tables - we simply highlight R1CS as a prevalent constraint system. The Groth16 zk-SNARK generation and verification process can be represented at a high-level with the following algorithms:
\begin{itemize}
    \item $(\mathcal{VK, PK})\xleftarrow[]{}$ Setup(\Cir): A trusted third party or \Vrf run a setup procedure to generate a prover key $\mathcal{PK}$ and verifier key $\mathcal{VK}$. These keys are used for proof generation and verification, respectively. This setup must be repeated each time \Cir changes.
    \item $\pi \xleftarrow[]{}$ Prove($\mathcal{PK}$, \Cir, $x$, $y$, $w$): \Prv generates proof $\pi$ to convince \Vrf that $w$ is a valid witness.
    \item $1/0 \xleftarrow[]{}$ Verify($\mathcal{VK}$, \Cir, $x$, $y$, $\pi$): \Vrf accepts or rejects proof $\pi$. Due to soundness property of zk-SNARKs, \Vrf cannot be convinced that $w$ is a valid witness by a cheating \Prv.
\end{itemize}

\noindent In Appendix \ref{sec:recursive}, we describe how zk-SNARKs can be extended to allow recursive construction and verification of proofs.

As we stated, one of the drawbacks of traditional pre-processing zk-SNARKs is their reliance on a trusted setup per circuit \Cir.
PLONKS, a subset of zk-SNARKs, are a class non-interactive ZK protocols that improve upon pre-processing zk-SNARKs by getting rid of the trusted setup per circuit \Cir, while adding a bit more arithmetic flexibility \cite{plonk}. PLONKs utilize the idea of a universal and updatable trusted setup, introduced in theory by \cite{cryptoeprint:2018/280} and in practice by \cite{cryptoeprint:2019/099}, in which a trusted setup procedure is done for circuits up to a certain size. 
Every circuit \Cir that fits within these size constraints can utilize the parameters generated by the universal trusted setup process. While PLONKs introduce a universal trusted setup, it comes at the cost of proof size and \Vrf runtime. PLONK proofs are normally $2-5\times$ the size of zk-SNARKs, and \Vrf runtime is marginally higher. It is important to note that, although PLONK proofs are larger than those of zk-SNARKs, proof size still remains in the KB range. The advantage that PLONKs have is that they are flexible in the commitment scheme they can use. By using the standard Kate commitments \cite{kate2010constant}, PLONKs become more zk-SNARK-like, as these commitments are based on ECC. FRI commitments \cite{ben2018fast}, which rely on Reed-Solomon codes and low-degree polynomials/testing for verifiers, can also be used to make PLONKs more zk-STARK-like. The type of commitment schemes allows developers to balance the tradeoff between performance and security assumptions. PLONK arithmetization is similar to that of zk-SNARKs, meaning that the resulting representation is a set of polynomials. To get there, PLONKs sets constraints for each gate (e.g. multiplication, addition) from the arithmetic circuit representation of the computation in the form of  Lagrange polynomials. Once the constraints are set, a special permutation function is used to check consistency between commitments. Finally, a final set of polynomials is constructed to fully represent the given computation. Overall, PLONKs provide a method to flexibly construct ZKPs with a less stringent trusted setup requirement, at the slight cost of performance. 

\textbf{Zero-Knowledge Scalable Transparent Arguments of Knowledge (zk-STARKs)}, which can be thought of as interactive oracle proof (IOP)-based zk-SNARKS, completely remove the dependence on trusted setup. Rather than using randomness from a trusted party, these protocols use publicly verifiable randomness for generating the necessary parameters for proof generation and verification. zk-STARKs achieve post-quantum security guarantees by utilizing collision-resistant hash functions as their underlying cryptography, rather than ECC. This increased security comes at a cost, as zk-STARK proofs are typically an order of magnitude larger than zk-SNARKs and PLONKs, and require more computational resources to generate and verify \cite{ben2018scalable}. The main contributor towards these drawbacks are the underlying data structure that are used in proof generation: Merkle trees. In zk-STARKs, Merkle trees are used to create a compact representation of the computation's execution trace. During proof generation, the computation's execution trace is arithmetized into polynomials, which are verified by performing low-degree testing, a process which ensures that the polynomials are of expected degree. Low-degree testing is enabled by the use of FRI commitments \cite{habock2022summary}. The polynomials are evaluated at certain points to verify their correct represenation of the execution trace, and these evaluations are used as the leaf nodes of the Merkle tree. The root of the Merkle tree then acts as a sort of commitment to these evaluated polynomials, hence allowing the verifier to simply verify the root, rather than verifying the whole computation trace. \cite{ashur2018marvellous} The use of Merkle trees are what enable the \textit{scalability} of zk-STARKs. While the Merkle trees support efficient verification, the proof size is drastically increased due to the inclusion of the material needed for verification, such as the Merkle root, polynomial evaluations, FRI commitments, and necessary Merkle branches. We note that there are IOP-based zk-SNARKs that stray away from this general protocol, but these steps are the most consistently utilized in current literature. Overall, zk-STARKs primarily benefit from being scalable and post-quantum secure with no trusted setup, at a significant cost to proof size and \Prv/\Vrf computation.

\textbf{MPC-in-the-Head (MPCitH)} ZKPs are a class of ZK protocols that take a completely novel approach towards proof generation and verification. The primary cryptographic basis is secure multiparty computation (MPC). MPC is a cryptographic primitive that allows for $n$ parties to jointly compute a function $f(x_1, ..., x_n)$, on private inputs from each party, without leaking any information about the private inputs. One of the prominent approaches to enable MPC is secret sharing, in which parties distributes secret shares of their private inputs amongst each other to compute a function. MPCitH, proposed by \cite{ishai2007zero}, allows for \Prv to simulate the $n$ MPC parties and following computation locally, or "in the head". Theoretically, any MPC protocol that can compute arbitrary functions can be transformed into a MPCitH ZKP. For $n$ parties $\{P_1,...,P_n\}$, secret shares are generated by each party and distributed to every other party. For the underlying arithmetic, the circuit \Cir is defined in an MPC manner to operate on secret shared data. \Prv can then simulate each parties' computation of the circuits with the secret shares they obtained from all other parties. After this is complete, \Prv has $n$ sets of messages and data that were generated and received by each party, called views. \Prv uses a standard commitment scheme to generate $n$ view commitments. Finally, \Prv and \Vrf interactively verify a subset of these views for consistency and correctness \cite{sidorenco2021formal}. While MPCitH protocols are innately interactive, they can be made non-interactive using the Fiat-Shamir transform. Theoretically, a huge advantage of the MPCitH approach is that MPC-friendly optimizations, which have been much further studied, can be utilized during proof generation to drastically improve \Prv efficiency and proof length. However, the most effective optimizations for MPC may translate to effective solutions for MPCitH. One of the core parameters MPCitH schemes aim to minimize is the communication complexity, as this directly reduces the amount of data that is present in each party's committed view. Just like zk-STARKs, MPCitH-based ZKPs do not require a trusted setup and are post-quantum secure, as MPC is thought to be generally quantum secure \cite{sidorenco2021formal}. Overall, MPCitH proposes a unique approach towards ZKP construction that are transparent post-quantum secure that allows flexibility in the underlying arithmetic to optimize the cost of proof generation and verification and proof size.

\textbf{Vector Oblivious Linear Evaluation (VOLE)-based ZK} 
protocols are a set of interactive techniques that achieve high efficiency and scalability through the use of information-theoretic message authentication code (IT-MAC)-based commitment schemes, which can be efficiently implemented using VOLE correlations. In VOLE-based ZKP protocols, the prover acts as the VOLE sender, while the verifier takes on the role of the VOLE receiver.
VOLE correlations are a pair of random variables, (\textbf{u}, \textbf{x}), known by \Prv and (\textbf{v}, $\Delta$), only known by \Vrf, in which \textbf{u}, \textbf{x}, and \textbf{v} are vectors, and $\Delta$ is a scalar key \cite{}. These variables satisfy the relation:
\begin{equation*}
    u_i = v_i + x_i \cdot \Delta
\end{equation*}
This functionality typically operates over a finite field.
Generally, in VOLE-based ZK, IT-MACs are used as commitments to authenticated wire values in arithmetic or boolean circuits representing a computation $\mathcal{C}$. \Prv demonstrates knowledge of a private vector $\textbf{w}$, which represents the witness, where $\mathcal{C}(\textbf{w}) = 1$, while proving the consistency throughout the protocol, without revealing any information about \textbf{w}.
\nojan{review and add VOLEitH}
VOLE-based proofs provide unparalleled scalability and communication optimizations, however, they are inherently designated-verifier protocols, meaning that \Prv must communicate with every \Vrf that aims to verify the proof, as \Vrf must maintain the secret $\Delta$ to ensure soundness. To address this, \cite{baum2023publicly} proposes a new VOLE-based paradigm, entitled VOLE-in-the-head (VOLEitH), which enables non-interactive VOLE-based ZK.









\section{ZKP Libraries} 
\label{sec:libraries}
\nojan{Add swanky and picozk}
\begin{table*}[t!]
   \small
   \centering\resizebox{\textwidth}{!}{
   \begin{tabular}{cccccccc}
   \toprule
   & \multicolumn{3}{c}{\textbf{Usability}} & \multicolumn{3}{c}{\textbf{Accessibility}} & \\
   \cmidrule(lr){2-4}
   \cmidrule(lr){5-7}
   \textbf{Framework}&
   \textbf{Language(s)} & \textbf{Custom \Cir} & \textbf{License} & \textbf{Examples} & \textbf{Documentation} & \textbf{GitHub Issues} & \textbf{Last Major Update} \\ 
   
   \midrule
    & \multicolumn{7}{c}{\textbf{zk-SNARKs}} \\
   \midrule
   
   \textbf{Arkworks \cite{arkworks}} & Rust & \fullcirc & MIT, Apache-2 & \fullcirc & \cmark & \halfcirc & Dec. 2023 \\
   \textbf{Gnark \cite{gnark-v0.9.0}} & Go & \fullcirc & Apache-2 & \fullcirc & \cmark & \fullcirc & Dec. 2024 \\
   \textbf{Hyrax \cite{hyraxZK}} & Python & \halfcirc & Apache-2 & \halfcirc & \xmark & \emptycirc & Feb. 2018 \\
   \textbf{LEGOSnark \cite{legosnark}} & C++ & \halfcirc & MIT, Apache-2 & \halfcirc & \xmark & \emptycirc & Oct. 2020 \\
   \textbf{LibSNARK \cite{libsnark}} & C++, Java (xJsnark \cite{kosba2018xjsnark}) & \fullcirc & MIT & \fullcirc & \xmark & \halfcirc & Jul. 2020 \\
  \textbf{Zokrates \cite{eberhardt2018zokrates}} & Zokrates DSL & \fullcirc & LGPL-3.0 & \fullcirc & \cmark & \halfcirc & Nov.2023 \\
   \textbf{Mirage \cite{Mirage}} & Java & \halfcirc & MIT & \halfcirc & \xmark & \emptycirc & Jan. 2021 \\
   \textbf{PySNARK \cite{PySNARK}} & Python & \fullcirc & Custom (MIT-like) & \fullcirc & \cmark & \halfcirc & May 2023 \\
   \textbf{SnarkJS \cite{baylina2020iden3}} & Circom \cite{munoz2022circom} & \fullcirc & GPL-3 & \fullcirc & \xmark & \halfcirc & Oct. 2024 \\
   \textbf{Rapidsnark \cite{RapidSNARK}} & Circom \cite{munoz2022circom} & \fullcirc & GPL-3 & \fullcirc & \xmark & \emptycirc & Dec. 2023 \\
   \textbf{Spartan\cite{Spartan}} & Rust & \emptycirc & MIT & \halfcirc & \xmark & \halfcirc & Jan. 2023 \\
   \textbf{Aurora (libiop) \cite{SciprLab2023Libiop}} & C++ & \emptycirc & MIT & \halfcirc & \xmark & \halfcirc & May 2021 \\
   \textbf{Fractal (libiop) \cite{SciprLab2023Libiop}} & C++ & \emptycirc & MIT & \halfcirc & \xmark & \halfcirc & May 2021 \\
   \textbf{Virgo \cite{SunblazeUCB2023Virgo}} & Python & \emptycirc & Apache-2 & \halfcirc & \xmark & \emptycirc & Jul. 2021 \\
     \textbf{Noir \cite{Noir2023Documentation}} & Rust DSL & \fullcirc & MIT, Apache-2 & \fullcirc & \cmark & \fullcirc & Nov. 2024 \\
   \textbf{Dusk-PLONK \cite{DuskPlonk2023Rust}} & Rust & \emptycirc & MPL-2 & \halfcirc & \cmark & \fullcirc & Aug. 2024 \\
   \textbf{Halo2 \cite{Halo22023Book}} & Rust & \halfcirc & MIT, Apache-2 & \fullcirc & \cmark & \fullcirc & Nov. 2023 \\

   \midrule
   & \multicolumn{7}{c}{\textbf{MPC-in-the-Head}} \\
   \midrule

   \textbf{Limbo \cite{KULeuvenCOSIC2023Limbo}} & Bristol \cite{bristol} & \fullcirc & MIT & \halfcirc & \xmark & \emptycirc & May 2021 \\
 \textbf{Ligero (libiop) \cite{SciprLab2023Libiop}} & C++ & \emptycirc & MIT & \halfcirc & \xmark & \halfcirc & May 2021 \\ 

   \midrule
   & \multicolumn{7}{c}{\textbf{VOLE-Based ZK}} \\
   \midrule
   \textbf{Mozzarella \cite{baum2022moz}} & Rust & \emptycirc & MIT & \halfcirc & \xmark & \emptycirc & Mar. 2022 \\
   \textbf{Diet Mac'n'Cheese \cite{dietmc}} & PicoZK \cite{picozk} & \fullcirc & MIT & \fullcirc & \xmark & \emptycirc & Sep. 2024 \\
  \textbf{Emp-ZK \cite{empzk}} & C++ & \fullcirc & MIT & \fullcirc & \xmark & \halfcirc & Sep. 2023 \\

   \midrule
   & \multicolumn{7}{c}{\textbf{zk-STARKs}} \\
   \midrule

   \textbf{MidenVM \cite{PolygonMiden2023MidenVM}} & Miden Assembly & \halfcirc & MIT & \halfcirc & \cmark & \fullcirc & Nov. 2023 \\
   \textbf{Zilch \cite{TrustworthyComputing2023Zilch}} & Java DSL & \halfcirc & MIT & \fullcirc & \xmark & \emptycirc & Apr. 2022 \\
   \textbf{RISC Zero \cite{RISCZero2023DeveloperDocs}} & Rust, C++ & \fullcirc & Apache-2 & \fullcirc & \cmark & \fullcirc & Dec. 2024 \\
   
   \bottomrule
   \end{tabular}}
   \caption{ZK Framework Attributes} 
      \label{tab:usability}
\end{table*}

In this section, we discuss the details of the 25 frameworks that we target in this work. We aim to highlight frameworks bred from both industry and academia. We primarily focus on works that present novel implementations of proving schemes that can be integrated with their own exposed high-level API for custom circuit design, or a general-purpose ZKP circuit development frontend, such as Circom \cite{munoz2022circom} or Zokrates \cite{eberhardt2018zokrates}.

Alongside the in-depth descriptions of each framework, we provide an evaluation of these frameworks at a high-level on usability and accessibility metrics, presented in Table \ref{tab:usability}. Our measurement of some metrics require further explanation:
\begin{itemize}
    \item Custom \Cir: \fullcirc $=$ Non-cryptography software engineer can build custom circuits, \halfcirc $=$ Building custom circuits requires deep knowledge of syntax; A developer could not read the code and understand it, \emptycirc $=$ Custom circuits require deep knowledge of protocol and syntax; normally requires manual translation of constraints to gates
    \item Examples: \fullcirc$=$ Plenty of examples are shared that fully show the capabilities of the system, \halfcirc $=$ Examples are included, but are not representative of the system's full capabilities
    \item Github Issues: \fullcirc $=$ Users and developers are both active in issues forum, \halfcirc $=$ Users are relatively active and developers are sporadically active, \emptycirc $=$ No activity
\end{itemize}

Table \ref{tab:description} in Appendix \ref{sec:app_libraries} outlines the discussed frameworks at a high level.




\subsection{zk-SNARKs}

\textbf{libsnark.} The \texttt{libsnark} C++ development library \cite{libsnark} is widely regarded as the original and most well-developed library for zk-SNARKs. This is highlighted by the fact that Zcash, the first real-world implementation of zk-SNARKs, was built upon \texttt{libsnark}. \texttt{libsnark} supports the Pinocchio \cite{parno2016pinocchio} and Groth16 \cite{groth16} proving schemes, alongside many different underlying elliptic curves. Much of the novelty of \texttt{libsnark} comes from the different forms of circuits that it supports. It supports R1CS and QAPs, as most frameworks do, but also supports higher level forms such as Unitary-Square Constraint Systems (USCS) and Two-input Boolean Circuit Satisfiability (TBCS) \cite{uscs}. The scheme used in \texttt{libsnark} is described as a preprocessing zk-SNARK, which simply highlights that trusted setup is performed before proof generation and verification. \texttt{libsnark} provides low-level "gadgets", which can be combined and built upon to represent the desired computation in R1CS format, however it is not the easiest way to develop zk-SNARKs in this library. \cite{kosba2018xjsnark} presents \texttt{xJsnark}, a high-level Java framework that allows a user to essentially code their computation in standard Java. Behind the scenes, this framework optimizes computation and outputs the computation in R1CS format. This output can be used directly with \texttt{libsnark}'s zk-SNARK generation script. Combined with \texttt{xJsnark}, \texttt{libsnark} is a highly-accessible option for inexperienced ZKP developers.

\textbf{gnark.} The \texttt{gnark} library \cite{gnark-v0.9.0} enables developers to build zk-SNARK-based applications using the high-level API it offers in Go language. The primary focus of \texttt{gnark} is runtime speed \cite{ConsenSys2023Gnark}. It offers both Groth16~\cite{groth16} and PLONK~\cite{plonk} (with KZG and FRI polynomial commitment) SNARK protocols. It offers a lot of curves, and can build R1CS circuits. In terms of hashing, it offers MiMC~\cite{mimchash}, SHA2, and SHA3 gadgets out-of-the-box. It also offers a collection of high-level gadgets for ease of building custom circuits. This framework exposes a high-level API that allows users to build their own gadgets, while utilizing the Go standard language and the provided gadgets. Recently, \texttt{gnark} has introduced GPU support with the support of the Icicle library \cite{icicle}. This work is in active development and seems to have an active community around it, making it an accessible option for inexperienced ZKP developers. We recommend this for beginners and experts alike for almost any custom applications. This framework utilizes a readable and robust API that any user can take advantage of and build custom applications with.

\textbf{arkworks.} The \texttt{arkworks} Rust ecosystem~\cite{arkworks} is an extensive and modular collection of libraries that can be used for efficient zk-SNARK programming. This ecosystem provides highly efficient implementations of arithmetic over \textit{various} curves and fields, even allowing curve specific optimizations. The main offering of \texttt{arkworks} is a generic application development framework that supports both experienced and non-experienced zk-SNARK developers. This framework enables high-level zk-SNARK development, as it allows users to implement their circuit as constraints (R1CS), while abstracting out details of SNARKs and curves, using an \texttt{arkworks} library. To venture into lower-level optimizations, \texttt{arkworks} provides libraries for the user to describe their circuit in native code. This allows the users to make several design decisions, such as specifying which proving system, such as Groth16, they would like to use. Alongside this, \texttt{arkworks} also provides libraries implementing low-level finite field, elliptic curve, and polynomial interfaces. In addition to SHA256, ZKP-friendly hashes such as Pedersen \cite{pedersenhash} and Poseidon \cite{poseidonhash} hashing are also offered. The \texttt{arkworks} development ecosystem is actively maintained and has an active community. We recommend this framework for users that have a deep knowledge of ZKPs, as one of the main advantages of arkworks, other than it's fantastic and usable codebase, is the ability to tweak certain parameters to optimize operations for your custom application.

\textbf{hyraxZK.} \label{sec:hyrax} \texttt{Hyrax} is a "doubly-efficient" zk-SNARK scheme, providing a concretely efficient prover and verifier, with low communication cost and no trusted setup \cite{wahby2018doubly}. Instead of following a standard underlying zk-SNARK structure, \texttt{Hyrax} is built on top of the Giraffe interactive proof scheme \cite{wahby2017full}. The authors apply a technique to reduce communication cost and add cryptopgraphic operations to turn the interactive proof into a ZKP. With the addition of optimized cryptographic commitments, the concrete cost of this scheme is significantly reduced and results in an interactive ZKP scheme. Using the Fiat-Shamir transform \cite{kilian1992note}, this scheme is made non-interactive. \texttt{hyraxZK} \cite{hyraxZK} provides a cleanly-developed Python and C++ development environment using \texttt{Hyrax} as the underlying zk-SNARK scheme. The provided framework is well-developed, however there is a lack of documentation that makes it challenging to build custom circuits.

\textbf{libspartan.} \texttt{libspartan} \cite{Spartan} is a Rust library that implements the \texttt{Spartan} zk-SNARK proof system \cite{setty2020spartan}. \texttt{Spartan} is a transparent zk-SNARK proof system, meaning that it requires no trusted setup. \texttt{libspartan} utilizes a Rust implemention of group operations on prime-order group Ristretto \cite{ristretto} and elliptic curve Curve25519 \cite{bernstein2006curve25519}, which ensures security and speed.
By adding a new commitment scheme, alongside a novel cryptographic compiler and a compact encoding of R1CS instances, \texttt{Spartan} is able to achieve the first transparent proof system with sub-linear verification costs and a time-optimal prover, at the cost of memory-heavy computation on the prover side. \texttt{libspartan} is a well-developed and maintained framework, however implementing custom functions is not very straightforward based on the provided documentation. Developing a custom ZKP circuit in \texttt{libspartan} requires the user to have the parameters of the R1CS instance, alongside knowledge of how to encode the constraints into R1CS matrices. Depending on the size of the ZKP circuit, this process can be very rigorous and involved, while also requiring a full knowledge of R1CS representations.
\texttt{Zokrates} \cite{eberhardt2018zokrates} provides a high-level API to build an R1CS for custom ZKP circuits, however a developer then has to manually convert these into a format that is readable by \texttt{libspartan}, which can be time-intensive depending on the number of constraints in the circuit. We only recommend this framework to users that have an in-depth knowledge of ZK constraint systems, however, we do note that this framework's backend is state-of-the-art and, upon integration with a standard frontend, would be a perfect solution for most ZK applications.

\textbf{Mirage.}
\texttt{Mirage} \cite{kosba2020mirage} is a universal zk-SNARK scheme and aptly named Java framework \cite{Mirage} implementing such scheme. \texttt{Mirage}'s main contribution is a universal trusted setup, such that trusted setup does not have to be performed everytime the circuit changes, as is done in zk-SNARKs. This saves a great amount of time and computation at the cost of higher proof computation overhead. This work introduces the idea of \textit{separated zk-SNARKs}, which enables efficient randomized checks in zk-SNARK circuits. This results in simplified verification complexity. Combining this with their novel universal circuit generator that produces circuits linear in the number of additions and multiplications, the \texttt{Mirage} zk-SNARK scheme is introduced. The underlying scheme and circuit generator are implemented in the \texttt{mirage} codebase, which has a Java frontend for circuit generation and a C++ backend implementing \texttt{Mirage} on top of \texttt{libsnark}. The core of development is done in \texttt{mirage}'s universal circuit generator, as that is where the ZKP circuits are specified by the user. This codebase provides very readable and diverse examples that highlight the use cases of their high-level Java API. 
Not only is there a bit of a learning curve to get acquainted with \texttt{mirage}'s syntax, but we also found that the codebase is relatively outdated, meaning that the code no longer compiles.

\textbf{LegoSNARK.} 
\texttt{LegoSNARK} \cite{campanelli2019legosnark} is a zk-SNARK scheme and library that focuses on linking SNARK "gadgets" together to build zk-SNARKs with a modular approach. This library implements the modular zk-SNARKs in the form of commit-and-prove zk-SNARKs (CP-SNARKs) \cite{lipmaa2016prover}, which are a class of zk-SNARKs that prove statements about committed values. As previous CP-SNARK schemes are limited due to their reliance on a single commitment scheme, one of the most important contributions of this work is a generic construction that can convert a broad class of zk-SNARKS, such as QAP-based, to CP-SNARKs. The \texttt{LegoSNARK} library \cite{legosnark} provides end-to-end proving and verification using the proposed scheme in a C++ package. This work builds upon \texttt{libsnark}, albeit with integration to high-level \texttt{libsnark} frameworks, such as \texttt{xJsnark}. Nevertheless, this library provides readable examples for developing gadgets, making it relatively easy for experienced C++ developers to build custom gadgets for their ZK applications without an in-depth knowledge of ZKPs. We recommend this framework to users that are building modular applications that benefit from CP-SNARKs, such as matrix arithmetic.


\textbf{PySNARK.}
\texttt{PySNARK} \cite{PySNARK} is a Python library that allows developers to use pure Python syntax to develop zk-SNARKs with various backends. PySNARK gives users access to \texttt{libsnark}, \texttt{qaptools}, \texttt{zkinterface}, and \texttt{snarkjs} backends. Compiling computation with the \texttt{libsnark} and \texttt{qaptools} performs proof generation and verification using the Groth16 and Pinnochio proving systems, respectively. Using the \texttt{zkinterface} backend simply generates \texttt{.zkif} files that can be used with the \texttt{zkinterface} package for proof generation and verification, where the underlying scheme can be chosen. Similarly, using the \texttt{snarkjs} backend generates the witness and R1CS files that can be used within our provided \texttt{snarkjs} environment. Overall, \texttt{PySNARK} is a brilliantly documented and developed library for beginners with zk-SNARKs, however it is not actively maintained. Developers that are comfortable with Python should have no trouble developing ZK applications once they become familiar with the library's syntax. Due to the Python compilation process, \texttt{PySNARK} experiences non-ideal operation times, so users should primarily use this for testing applications on the Groth16 proving system, but not for practical application development.

\textbf{SnarkJS + RapidSNARK.}
\texttt{SnarkJS} \cite{baylina2020iden3} is built on Javascript (JS) and Pure Web Assembly (WASM) and supports the Groth16, PLONK, and FFLONK underlying proving schemes. This framework accepts circuits designed in \texttt{circom} \cite{munoz2022circom}, which provides a very accessible frontend with a well-documented API for building ZK circuits. The protocols that are supported all require trusted setup, whether it be a circuit-specific setup for Groth16, or a universal setup for PLONK/FFLONK. Also, switching between ZK schemes is simply done by specifying the desired scheme as a command line argument. \texttt{SnarkJS} provides a multi-step universal setup protocol that all programs perform, alongside a Groth16-specific setup. Alongside this, the circuit to proof compilation process is done in a modular way that allows for closer debugging. In the proof generation process, the circuit characteristic's are listed for the developer (e.g. constraints, public inputs) which enables quick sanity checks. Finally, \texttt{SnarkJS} provides simple routes to turning the verifier into a smart contract, or performing the end-to-end ZKP process in browser, due to the JS and WASM backend. \texttt{RapidSNARK} \cite{RapidSNARK} is built upon C++ and Intel assembly by the same developers, and significantly improves upon \texttt{SnarkJS}. Using a very similar API, and even accepting \texttt{SnarkJS}-generated files as inputs (e.g. proving/verifier keys, witness), \texttt{RapidSNARK} allows for faster proof generation with a simple change in command line arguments from the \texttt{SnarkJS} commands. The main advantage of this framework is the utilization of parallelization within proof generation, yielding much faster results than \texttt{SnarkJS}, however the downside is that only Groth16 proofs are supported. While \texttt{SnarkJS} is more actively maintained than \texttt{RapidSNARK}, both frameworks are highly accessible for those with little experience in developing ZK applications, due to the ability to utilize a \texttt{circom} frontend.

\textbf{Virgo.}
\texttt{Virgo} \cite{zhang2020transparent} is an implementation of a novel interactive doubly-efficient ZK argument system. The main advantage of this protocol is the lack of trusted setup, which is oftentimes the most cumbersome task in zk-SNARKs. \texttt{Virgo} sees the most benefits for layered arithmetic circuits, rather than all general arithmetic circuits, as it is based off the GKR protocol \cite{goldwasser2015delegating}, which also is only catered towards structured circuits. General arithmetic circuits are addressed in a follow up work, \texttt{Virgo++} \cite{virgoplus}. The open-source implementation of this work does not have ZK commitments implemented yet, which is why we do not consider it in our survey. The main enabling factor of \texttt{Virgo} is a novel ZK verifiable polynomial delegation (zkVPD) scheme, which can essentially be seen as a commitment scheme in this scenario. 
Due to the reliance on zkVPD and the allowed interactivity in this scheme, the implementation only relies on lightweight cryptography, making it a feasible development solution. While an impressive solution with great results, the repository is not actively maintained and lacks clear documentation, meaning it is not the most suitable candidate for ZK application developers.

\textbf{libiop} \label{sec:aurora}
The \texttt{libiop} framework \cite{SciprLab2023Libiop} is a collection of three protocol implementations: Aurora \cite{aurora}, Fractal \cite{fractal}, and Ligero \cite{ames2017ligero}. Ligero falls under the MPCitH category, so it is discussed later in the paper. Aurora and Fractal are both post-quantum, transparent zk-SNARKs, which classifies them more as succinct zk-STARKs. However, the authors classify their work as zk-SNARKs, which is why they are discussed here. Both works outperform prior zk-SNARKs by proposing new interactive oracle proofs (IOPs). Fractal proposes a holographic IOP \cite{babai1993transparent}, while Aurora proposes an IOP based around Reed-Solomon codes.
As for the \texttt{libiop} implementations, it does not seem to be actively maintained. While there are a few example applications for each protocol, the most useful tool in was the benchmarking scripts that were provided. This allows users to input parameters, such as number of constraints and variables, to specify a random circuit and outputs the performance metrics of the protocol. This shows how the protocols scale based on the size of the circuit. These parameters can be extracted from R1CS files (made by frameworks such as Zokrates), using our provided \texttt{R1CSReader} scripts. While the benchmarking is convenient, developing custom applications with this framework requires a deeper knowledge of the protocol that may not be easily accessible to all developers. We only recommend this to users that have a deep knowledge of the literature that these frameworks stem from.

\textbf{Noir.}
\texttt{Noir} \cite{Noir2023Documentation} is a general Rust-like framework for developing applications based on ZKPs. Fundamentally, \texttt{Noir} is a domain-specific language that resembles Rust. It enables one to build circuits that implement complex logic without having to learn the low-level details of ZKP systems. Since it acts like a generalized front end, it is capable of building circuits for a variety of back ends. Currently, Barretenberg \cite{AztecProtocol2023Barretenberg} serves as the default back end, and generates PLONK proofs and Solidity contracts. The Barretenberg back end can also use WASM to create proofs and verify them directly in the browser. Arkworks is also available as an out-of-the-box back end, which can generate Groth16 and Marlin proofs. This generalization is possible because Noir framework compiles the circuit to an intermediate language referred to as ACIR (Abstract Circuit Intermediate Representation), which can then be further compiled to specific R1CS or arithmetic circuit compatible with a specific back end. The framework also provides a Typescript library for direct integration into web applications. There is active development going on, but Noir currently supports a full control flow with the ability to create custom circuits using readable code. This is a great option for developers who would like to avoid the details of ZKPs and build applications using a Rust-like DSL. We recommend this for those who want to build simplistic applications who have little experience with ZKPs.




\textbf{Dusk-PLONK.}
\texttt{Dusk-PLONK} \cite{DuskPlonk2023Rust} is a pure Rust implementation of the PLONK proving system. This implementation supports operation over the BLS12-381 and JubJub elliptic curves. The developers of this framework use Kate commitments \cite{kate2010constant} as their primary polynomial commitment scheme to utilize its homomorphism and maintain constant size commitments. The provided codebase is extremely detailed and well-commented and provides helpful documentation. Similar to other PLONK frameworks, \texttt{Dusk-PLONK} only provides a very low-level API for custom circuit development. To build a custom circuit, developers must translate their computation into an arithmetic or boolean circuit gate format (e.g. add, multiply). This is perfectly digestible for small circuits, as shown in the examples, however becomes an intensely laborious task as the circuit and number of inputs or input dimensions scales up. While the code is well-written and yields excellent results, this framework requires a more sophisticated high-level API that utilizes common software engineering structures to build custom circuits before new developers can start building practical ZKP applications with it. We do note that this is a fantastic implementation of the PLONK proving system for and recommend it for developers that have experience with logic design and ZKPs. 

\textbf{Halo2.}
Built by the same creators of Zcash and the original Halo \cite{bowe2019recursive} framework, the Halo2 framework \cite{Halo22023Book} optimizes upon some of the inefficiencies of its predecessors by utilizing a PLONK-ish scheme as the underlying proving system. The underlying polynomial commitment scheme in this framework is Kate commitments. In its original repository and documentation, building a custom circuit with Halo2 requires a developer to design their computation in the form of a circuit, by implementing gates and utilizing them to build a \textit{chip}. This can be relatively confusing for new developers.
However, Halo2 is a powerful proof system that is utilized widely across the industry, including a prominent verifiable machine learning framework, \texttt{ezkl} \cite{ZkonduitInc2023EZKL}. This prominence has garnered a strong community backing the framework and has resulted in many works that either provide more examples of how the framework can be used \cite{Halo2Club2023}, or expose higher-level APIs for building custom circuits. Overall, while the Halo2 framework only exposes a lower-level API for custom circuit building, the community around it makes it a relatively accessible solution for practical application of PLONKs. We believe this is a good framework for those experienced with applied cryptography and interest in building machine-learning focused applications.





\subsection{MPC-in-the-head}

\textbf{Ligero (libiop).}
The Ligero \cite{ames2017ligero} protocol is implemented in \texttt{libiop} \cite{SciprLab2023Libiop} framework. This interactive protocol applies the general IKOS \cite{ishai2007zero} transformation that transforms MPC-based interactive proofs into ZKPs, which is typical for MPC-in-the-Head (MPCitH) systems. This means that the key aspect of designing the Ligero is the underlying MPC protocol. While this protocol is interactive, it can be transformed into a zk-SNARK using the Fiat-Shamir transform, just like any other interactive protocol. Additionally, the Ligero protocol only relies on collision resistant hash functions for the underlying cryptography and does not require a trusted setup. As this is implemented using the same backend as the Aurora and Fractal zk-SNARK protocols, all implementation details remain the same as described in section \ref{sec:aurora}.

\textbf{Limbo.}
Similar to Ligero, \texttt{Limbo}'s implementation \cite{KULeuvenCOSIC2023Limbo}  and underlying protocol \cite{limbo} is reliant on the IKOS transformation that MPCitH protocols often rely on. \texttt{Limbo} improves upon Ligero by highlighting the tradeoff between MPCitH parties involved, proof size, and runtime.
The main work \texttt{Limbo} compares to is Ligero, as they are both transparent MPCitH schemes that only rely on collision resistant hash functions. \texttt{Limbo} claims to work better on small and medium circuits. While the \texttt{Limbo} framework is not as extensively developed, maintained, and documented as some of the other frameworks highlighted in this work, it greatly benefits from its ability to take Bristol Circuit (BC), a common way to describe MPC circuits \cite{bristol}, descriptions as inputs. This allows developers to build custom applications by describing their general computations in BC format. We provide a simple pipeline for developing BCs, alongside examples using readable syntax. We recommend this for users who have experience building optimized BCs and have a relatively deep understanding of MPC.



\subsection{VOLE-Based ZK}

\textbf{Diet Mac'n'Cheese}
\nojan{add swanky stuff}
\texttt{Diet Mac'n'Cheese} \cite{dietmc} is a novel framework that implements the Mac'n'Cheese protocol \cite{baum2021mac}, a Vector Oblivious Linear Evaluation (VOLE)-based zero-knowledge protocol over the $\mathbb{Z}_{2^k}$ ring. Similar to Moz$\mathbb{Z}_{2^k}$arella, this is a crucial step in making ZKPs more practical, as most real-world compute hardware operates on integer rings, and not finite fields. \texttt{Diet Mac'n'Cheese} makes many improvements to the state-of-the-art in VOLE-based ZK protocols by optimizing the underlying sVOLE subprotocol. This optimization yields significant performance improvements over prior VOLE protocols that operate over integer rings. The provided implementation comes in the form of a C++ package that directly implements the proposed scheme and uses the Swanky ecosystem \cite{swanky} for easy integration. This framework is still in its early stages of development and currently lacks extensive documentation and concrete examples, making it harder for new ZKP developers to use it. Alongside this, \texttt{Diet Mac'n'Cheese} currently only supports fixed-point integer operations. It exposes a low-level API that requires a developer to explicitly define all computations as arithmetic and boolean gates that are operated on using the framework's provided functions. However, a recent work has introduced a Python frontend with great documentation that can translate Python code into an intermediate representation that is recognized by the \texttt{Diet Mac'n'Cheese} framework. This frontend, entitled PicoZK \cite{picozk}, contains many examples and is even able to integrate with the popular numpy and pandas packages. PicoZK is a perfect pairing with \texttt{Diet Mac'n'Cheese} and allows for the development of simple applications. We recommend this framework to any developer that aims to build a scalable application that is conducive to a designated-verifier environment, such as federated or split learning. We do note that any floating point operations that are done with this framework must be converted to fixed-point.

\textbf{emp-zk.}
\nojan{shorten} The \texttt{emp-zk} development framework \cite{empzk} is a part of the \texttt{emp-toolkit} \cite{emptool}, a collection of cryptographic front-ends and back-ends that allow for easy development of multi-party computation applications. Alongside ZKPs, \texttt{emp-toolkit} also provides libraries for garbled circuits and oblivious transfer. \texttt{emp-zk} has implementations of three novel interactive ZK systems:
\begin{itemize}
    \item Wolverine \cite{weng2021wolverine}, the first of these systems, presents a constant-round, scalable, and prover-efficient interactive ZK scheme.
    \item Mystique \cite{weng2021mystique}, built on top of Wolverine, focuses on machine learning applications. This work presents efficient conversions for arithmetic and boolean values, fixed-point and floating-point values, and committed and authenticated values. 
    \item Quicksilver \cite{yang2021quicksilver}, also built on top of Wolverine, further improves communication costs and scalability.
\end{itemize}

The main primitive these schemes take advantage of is subfield Vector Oblivious Linear Evaluation (sVOLE), which the authors extend and optimize for their ZK scheme. 
For sake of brevity, we spare the technical detail in this paper and refer to \cite{Weng2023VOLEBasedInteractive} for an excellent explanation. \texttt{emp-zk} provides a very user-friendly interface to all 3 ZK systems, with clear-cut examples. Although documentation is not explicitly provided, \texttt{emp-zk} largely relies on C++ syntax and does not require much knowledge about the underlying work in ZKPs, making it one of the more accessible options. One potential downside of these systems are that they are interactive, meaning all proofs are \textit{designated-verifier}. We highly recommend this framework for users who are building custom machine learning-based custom applications that rely on floating-point operations, or applications that rely on scalability (e.g. database operations).

\textbf{Moz$\mathbb{Z}_{2^k}$arella.}
This work \cite{baum2022moz} presents a new protocol that utilizes an novel vector oblivious linear evaluation (VOLE), a tool from secure two-party computation, extension to perform zero knowledge proof operations efficiently over the integer ring $\mathbb{Z}_{2^k}$. This is very important as most ZK systems are made to operate over finite fields, which is not representative of modern CPUs. The proof system is coined with the term \texttt{Quarksilver}. This protocol outperforms the previous state-of-the-art VOLE-based works that operate over finite fields. The accompanying implementation enables development of ZK applications with the \texttt{Quarksilver} protocol as the underlying scheme. The \texttt{Moz$\mathbb{Z}_{2^k}$arella} repository is not actively maintained, however has 3 sub-libraries for oblivious transfer, garbled and arithmetic circuits, and private set-intersection. Within these sub-libraries there are several examples that explain how to use the \texttt{Moz$\mathbb{Z}_{2^k}$arella} syntax, including examples for \texttt{Quarksilver}. While the examples are somewhat clear, using this library to build custom applications requires a deep knowledge of the underlying proof system, as users must be aware of the parameters that are being set on a per application basis. We only recommend this to users who's applications fully rely on using the specific underlying protocol in this framework.

\subsection{zk-STARKs}

\textbf{Miden VM.}
\texttt{Miden VM} \cite{PolygonMiden2023MidenVM} is a zero-knowledge virtual machine (zkVM) implemented in Rust, in which all programs that are run generate a zk-STARK that can be verified by anyone.
\texttt{Miden VM} is designed as a stack machine, consisting of a stack, memory, chiplets, and a host. The stack, the main user-facing component, is a push-down stack of field elements, which is where inputs and outputs of operations are stored. Increasing the amount of inputs that are initialized on the stack before program execution increases the verifier cost. Whatever is left on the stack after program computation is declared as a public input to the verifier, which also increases cost to the verifier. A prover's private inputs must be pushed to the stack during program computation to be kept private.
The aim of Miden VM is, in their own words, to "make Miden VM an easy compilation for high-level languages such as Rust" \cite{PolygonMiden2023VMOverview}. As these compilers do not yet exist, the only way to build custom circuits is using Miden's assembly language, a very low-level API that interfaces with the Miden stack, and Miden chiplets, which are optimized assembly-based modules that perform common operations, like field arithmetic. Although \texttt{Miden VM} is Turing complete and offers standard control flow, it is often challenging for a developer to translate their desired computation to assembly commands and managing the stack at the same time, especially as the size of computation scales up. While \texttt{Miden VM} is a very valuable tool, we believe that its highest potential will be achieved upon completion of an accompanying compiler from a high-level language to Miden assembly. We recommend that users use this to benchmark certain atomic operations, but to avoid building custom applications with this framework due to the lack of a frontend.


\textbf{Zilch.}
The \texttt{Zilch} framework \cite{mouris2021zilch} consists of a Java-like frontend (ZeroJava) that interfaces with a novel zero-knowledge MIPS processor model (zMIPS) \cite{TrustworthyComputing2023Zilch} to enable efficient interactive zk-STARK proof generation for custom computations. The ZeroJava frontend is highly sophisticated and is one of the only frameworks to enable an object-oriented programming approach. All ZeroJava programs are compiled into optimized and verifiable zMIPS instructions. As all of the instructions are verifiable, any program that can be expressed in ZeroJava can be verified using ZKPs. The underlying zMIPS processor can implement and verify any arbitrary computation in zero-knowledge. The zMIPS instructions are implemented using the zk-STARK library \cite{ben2018scalable}. After computation description in ZeroJava and compilation to zMIPS, the constraints for the program are represent in algebraic intermediate representation (AIR) format. The prover and verifier interactively undergo the zk-STARK process until the verifier is convinced that the prover's work is sound. 
\texttt{Zilch} provides an elegant and accessible approach to building custom circuits that utilize zk-STARKs. Although the works lacks dedicated documentation, the examples that are provided show that development of custom applications is almost as simple as implementing the program in Java, with a few ZeroJava design considerations. We recommend this for users with general knowledge of the MIPS instruction set architecture, which should allow them to build optimized programs.

\textbf{RISC Zero.}
\texttt{RISC Zero} is a zkVM \cite{RISCZero2023DeveloperDocs} implemented in Rust with an underlying RISC-V processor and instruction set architecture. The goal of this work is to produce publicly verifiable proofs of all the computations that are done within the framework. As the underlying instructions are derived from RISC-V, virtually any arbitrary computation can be expressed and verified in zero-knowledge.
In this framework, custom circuits can be built using standard Rust syntax, with a few minor modifications to incorporates the framework's API. This program is compiled to a set of RISC-V instructions, which is then executed within a \texttt{RISC Zero} session, which is recorded. A receipt of this session is recorded and used as part of the zk-STARK proof, which can be verified by any verifier to check validity of the computation.
\texttt{RISC Zero} provides a relatively readable high-level Rust API, alongside several examples and very detailed documentation. Due to the maturity of the Rust development and \texttt{RISC Zero} as a whole, developers are able to import a majority of the most used standard Rust crates without trouble, enabling much more streamlined and efficient application development. For instance, developers can use the JPG crate \cite{RustImageCrate2023} to build zero-knowledge applications around images. Alongside this, \texttt{RISC Zero} enables GPU acceleration, so that relevant applications can take advantage of computational speedup. We do note that although GPU acceleration is implemented in the RISC Zero codebase, we were not able to get it actually working due to some inconsistencies within the codebase. However, \texttt{RISC Zero} has an active community around it, including active development by the creators, and a very well-documented and accessible code, making it a great candidate for new developers of custom ZKP applications. The primary drawback for this framework is that, due the nature of zkVMs and the simulation of a RISC-V processor and ISA, this framework has relatively significant initialization and operation costs.
\section{Experimental Evaluation}
\begin{table*}[htb]
   \small
   \renewcommand{\arraystretch}{1.25}
   \centering\resizebox{\textwidth}{!}{
   \begin{tabular}{ccccc|cccc}
   \toprule
   & \multicolumn{4}{c}{\textbf{Matrix Multiplication}} & \multicolumn{4}{c}{\textbf{SHA-256}} \\

   \textbf{Framework} &
   \textbf{Setup (ms)} & \textbf{Prover (ms)} & \textbf{Comm./Proof Size} & \textbf{Verifier (ms)} & \textbf{Setup (ms)} & \textbf{Prover (ms)} & \textbf{Comm./Proof Size} & \textbf{Verifier (ms)} \\ 
   
   \midrule
    & \multicolumn{8}{c}{\textbf{zk-SNARKs}} \\
   \midrule
   
   \textbf{Arkworks (Groth16) \cite{arkworks}} & 31.939 & 45.665 & 128 B & 2.553 & 334.562 & 566.634 & 128 B & 1.310 \\
   \textbf{Arkworks (Marlin) \cite{arkworks}} & 423.386 & 403.624 & 951 B & 25.070 & \multicolumn{4}{c}{Unsupported operands} \\
   \textbf{Gnark (Groth16) \cite{gnark-v0.9.0}} & 182.896 & 37.449 & 164 B & 1.848 & 1154.924 & 149.497 & 164 B & 1.447 \\
  \textbf{Gnark (PLONK-FRI) \cite{gnark-v0.9.0}} & 1291.463 & 1444.085 & -\footnote{We could not measure proof size for this framework} & 2.594 & 135458.192 & 145301.453 & -$^3$ & 5.252 \\
   \textbf{Gnark (PLONK-KZG) \cite{gnark-v0.9.0}} & 47.396 & 21.638 & 552 B & 2.554 & 2806.739 & 635.682 & 552 B & 2.018 \\
  \textbf{Zokrates (Groth16) \cite{arkworks}} & 609 & 622 & 128 B & 310 & 1265 & 1296 & 128 B & 190 \\
    \textbf{Zokrates (GM17) \cite{arkworks}} & 782 & 807 & 96 B & 240 & 1411 & 1465 & 96 B & 180 \\
   \textbf{Hyrax \cite{hyraxZK}} & - & 4687.244 & 315 B & 317.408 & - & 5497.327 & 59.904 KB & 484.598 \\
   \textbf{LibSNARK \cite{libsnark}} & 160.3 & 179.547 & 127.375 B & 0.895 & 1579.5 & 588.2 & 127.375 B & 0.9 \\
   \textbf{PySNARK \cite{PySNARK}} & 1781.331 & 266.899 & 127.375 B & 4.561 & 31809.606 & 8006.642 & 127.375 B & 4.667 \\
   \textbf{SnarkJS (Groth16) \cite{baylina2020iden3}} & 3113 & 1410 & 802 B & 804 & 29100 & 1919 & 805 B & 637 \\
      \textbf{SnarkJS (PLONK) \cite{baylina2020iden3}} & 190632 & 282897 & 2.247 KB & 686 & 205550 & 378833 & 2.245 KB & 670 \\
  \textbf{Noir \cite{Noir2023Documentation}} & \multicolumn{2}{c}{6972.139\footnote{}} & 2.368 KB & 6037.378 & \multicolumn{2}{c}{10508.343$^2$} & 2.368 KB & 1154.979 \\
    \midrule
    & \multicolumn{8}{c}{\textbf{VOLE-based ZK}} \\
    \midrule
    \textbf{Emp-ZK \cite{empzk}} & 596.118 & 1.917 & 595.004 KB & 16.483 & 522.763 & 90.302 & 212.709 KB & 38.112 \\
    \textbf{Diet Mac'n'Cheese \cite{dietmc}} & 3817.626 & 4411.310 & 7.005 MB & 2397.265 & 3754.559 & 4861.536 & 3.558 MB & 4863.095 \\

   \midrule
    & \multicolumn{8}{c}{\textbf{MPC-in-the-Head}} \\
   \midrule
   \textbf{Limbo \cite{KULeuvenCOSIC2023Limbo}} & - & 96690.593 & 7.617802 MB & 72999.073 & - & 1129.368 & 113.57 KB & 879.399 \\
   \midrule

    & \multicolumn{8}{c}{\textbf{zk-STARKs}} \\
   \midrule

   \textbf{MidenVM \cite{PolygonMiden2023MidenVM}}& \multicolumn{4}{c|}{Memory overflow for this benchmark} & - & 514 & 71KB & 11 \\
   \textbf{RISC Zero \cite{RISCZero2023DeveloperDocs}} & - & 57268.609 & 279.640 KB & 59.058 & - & 4196.679 & 215.348 KB & 44.918 \\
   
   \bottomrule
   \end{tabular}}
   \caption{ZK Framework Performance. {\scriptsize $^2$ Noir only allows us to measure setup and prover time together. $^3$ PLONK-FRI did not allow for accurate proof measurement}}
   \label{tab:finalresults}
\end{table*}

\subsection{Configuration}
For all experiments, we build custom Docker environments that setup all dependencies and import all necessary programs to enable reproducible results. All reported results are the mean of 10 test runs. Benchmarking is done on a 128GB RAM, AMD Ryzen 3990X CPU desktop.

\subsection{Experimental Setup}
\nojan{Other than that, it would also have been nice to see some more comprehensive benchmarks that evaluate how these systems scale as the computation gets more complex; the reported benchmarks are for small circuits, which means that one-time initialization costs might dominate for some of the evaluated systems.}
This paper is focused on the usability and accessibility of ZKP frameworks and, more importantly, aims to serve as a guide to developers of novel ZKP-based applications. Our goal with this work is to allow new developers to have a full overview of the ZK development landscape after reading it. More importantly, we aim to provide a developer with the necessary insights to allow them to choose the framework that best suits their desired ZK-based application. Due to this developer-focused approach, we weed out some of the frameworks that we discuss in section \ref{sec:libraries}, due to the overhead that would be required for a new developer to build a custom circuit. This is not meant in a malicious manner to say the framework is not usable - these frameworks are state-of-the-art and provide excellent results. We simply are focused on frameworks that expose higher-level APIs, or can be easily integrated behind accessible frontends, for streamlined custom application development. Simply put, all the works discussed in section \ref{sec:libraries} are fantastic, and we highlight them as the best in the field. While some are missing a high-level API or frontend, they provide great value to the landscape. We still include their development environments with examples in our open-source repository to allow more experienced ZKP developers to easily access them and build applications with them.

While we recognize that there is no completely \textit{fair} way to benchmark these, we aim to do so by measuring the trusted setup (when applicable) and proof generation and verification runtimes, alongside communication for interactive protocols and proof size for non-interactive protocols. These are standard efficiency measures for ZKPs \cite{aurora}. While these quantitative results may not paint the whole picture, such as memory consumption and bandwidth considerations, we believe that they provide measurable proof of algorithmic complexity \nojan{edit} when independently benchmarked in the same isolated environment. We encourage readers to recreate our benchmarks, which are provided in our open-source repository, to gauge performance on their available hardware. To provide a rough estimate of the size of our benchmarks, we perform arithmetization to compile the benchmarks to R1CS format, and report the number of constraints of each circuit. Constraints are used often in ZK literature to describe the size of a ZK circuit. We evaluate the frameworks on the following benchmarks:


\nojan{A more principled approach towards comparing frameworks is needed: what should I use when I'm space-constrained? what should I use when I'm computationally constrained? what should I use if I have massive GPUs? The implementation-based comparison should answer all of these questions for this survey to elevate to the level of an SoK.}

\textbf{32$\times$32 Matrix Multiplication:} The prover aims to convince the verifier that they know two private matrices $\bm{\mathit{A^{32\times32}}}$ and $\bm{\mathit{B^{32\times32}}}$ that multiply to a public matrix $\bm{\mathit{C^{32\times32}}}$, without revealing anything about $\bm{\mathit{A}}$ or $\bm{\mathit{B}}$. This is a commonly used benchmark in this domain. This circuit is not too large and should be handled relatively easily by most frameworks, although some frameworks struggle with it due to memory issues. In R1CS format, this benchmark consists of 32,768 constraints.

\textbf{SHA-256:} \nojan{Find number of constraints for each app}
The prover aims to convince the verifier that they know $x$, a private 512-bit preimage, to the 2-to-1 hash function SHA-256$(x)=y$, where $y$ is a public 256-bit hashed value. This is quite a large circuit, compared to the matrix multiplication circuit. Some frameworks target this operation as one to optimize. We choose to evaluate on this benchmark as we believe it provides a good representation of a framework's performance, and it is a commonly used benchmark in this domain. In R1CS format, this benchmark consists of 59, 281 constraints.

\subsection{Results \& Takeaways} \nojan{add scalability tests to appendix}
\label{sec:results}
While we provide streamlined workflows for building custom applications for almost all frameworks that are discussed in section \ref{sec:libraries}, we narrow down our evaluation to frameworks that provide a novel, usable, and accessible approach for developing custom applications. Some frameworks, such as \texttt{Gnark} and \texttt{SnarkJS}, provide PLONK and zk-SNARK backends, so we evaluate our benchmarks on both backends. Overall, we analyze 18 systems for performance on our selected benchmarks. In this section, we discuss the implications of these results and the takeaways for developers. Alongside this, we provide recommendations for which frameworks or protocols are best to use in certain settings. The results can be seen in Table \ref{tab:finalresults}.

\begin{figure*}[t]
\centering
\subfloat[Scaling of trusted setup runtime.]{\includegraphics[width=.48\textwidth]{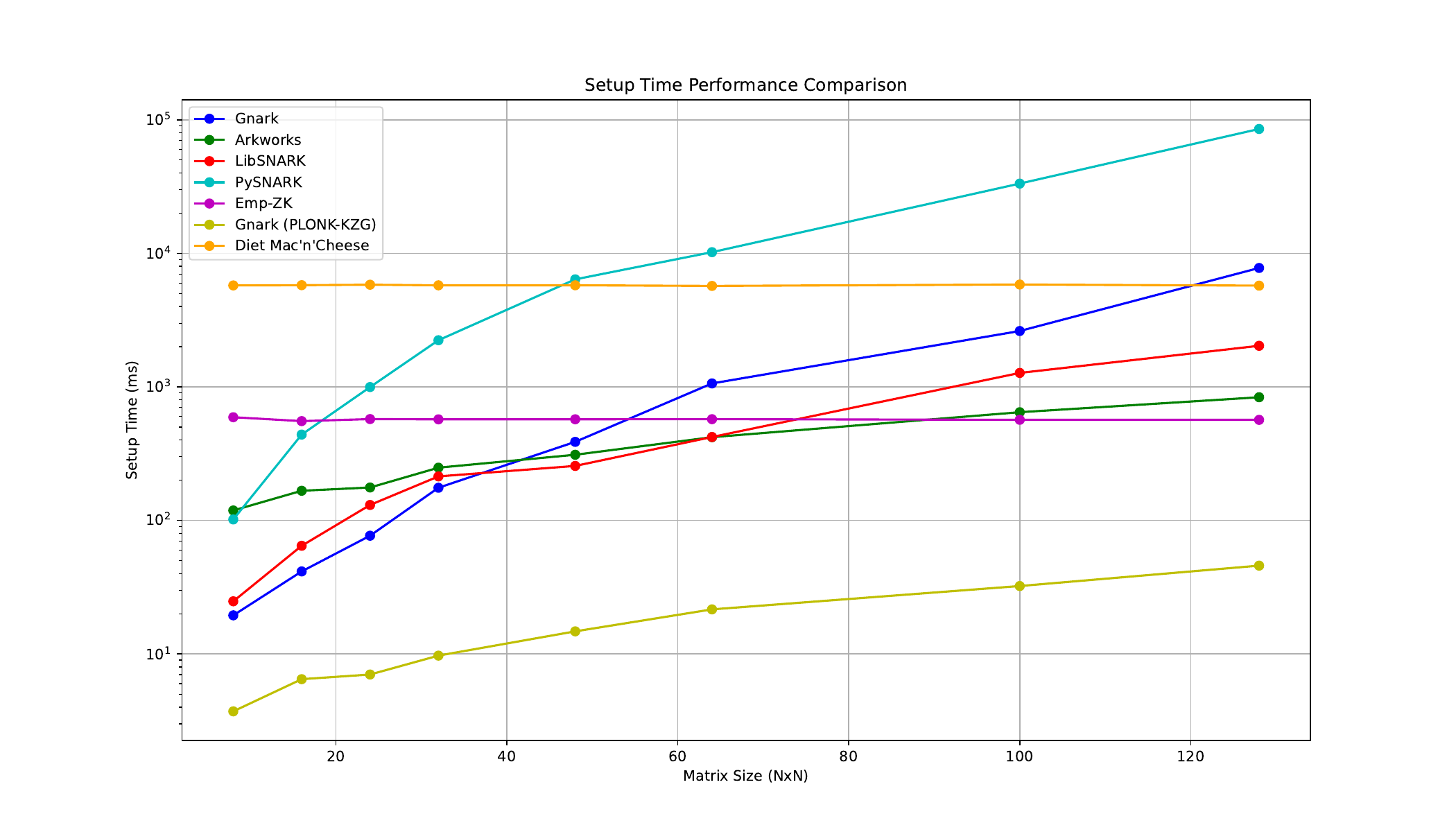}}
\hspace{1em}
\subfloat[Scaling of proof generation runtime.]{\includegraphics[width=.48\textwidth]{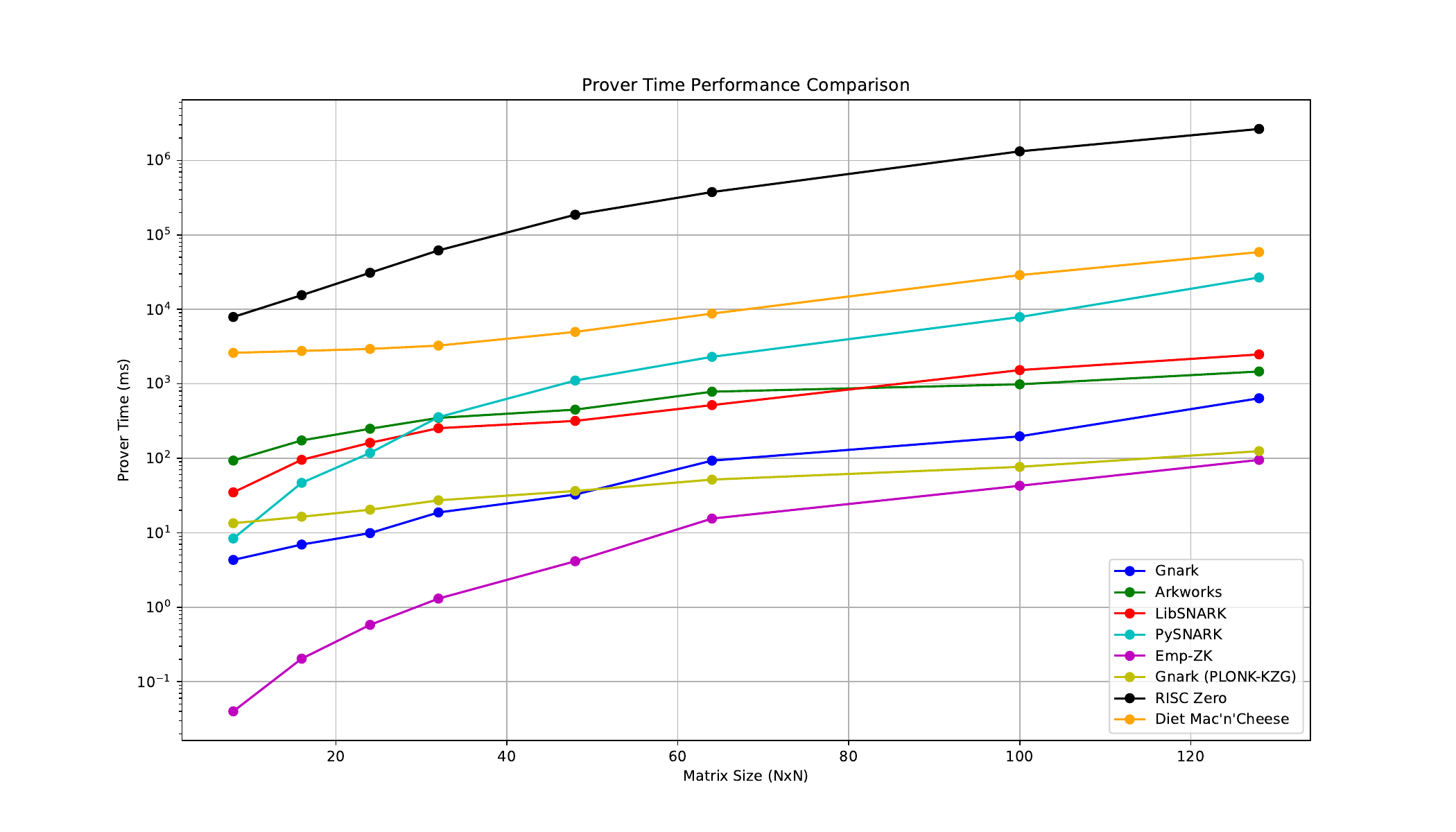}}
\hspace{1em}
\subfloat[Scaling of proof size (or communication in Emp-ZK's case).]{\includegraphics[width=.48\textwidth]{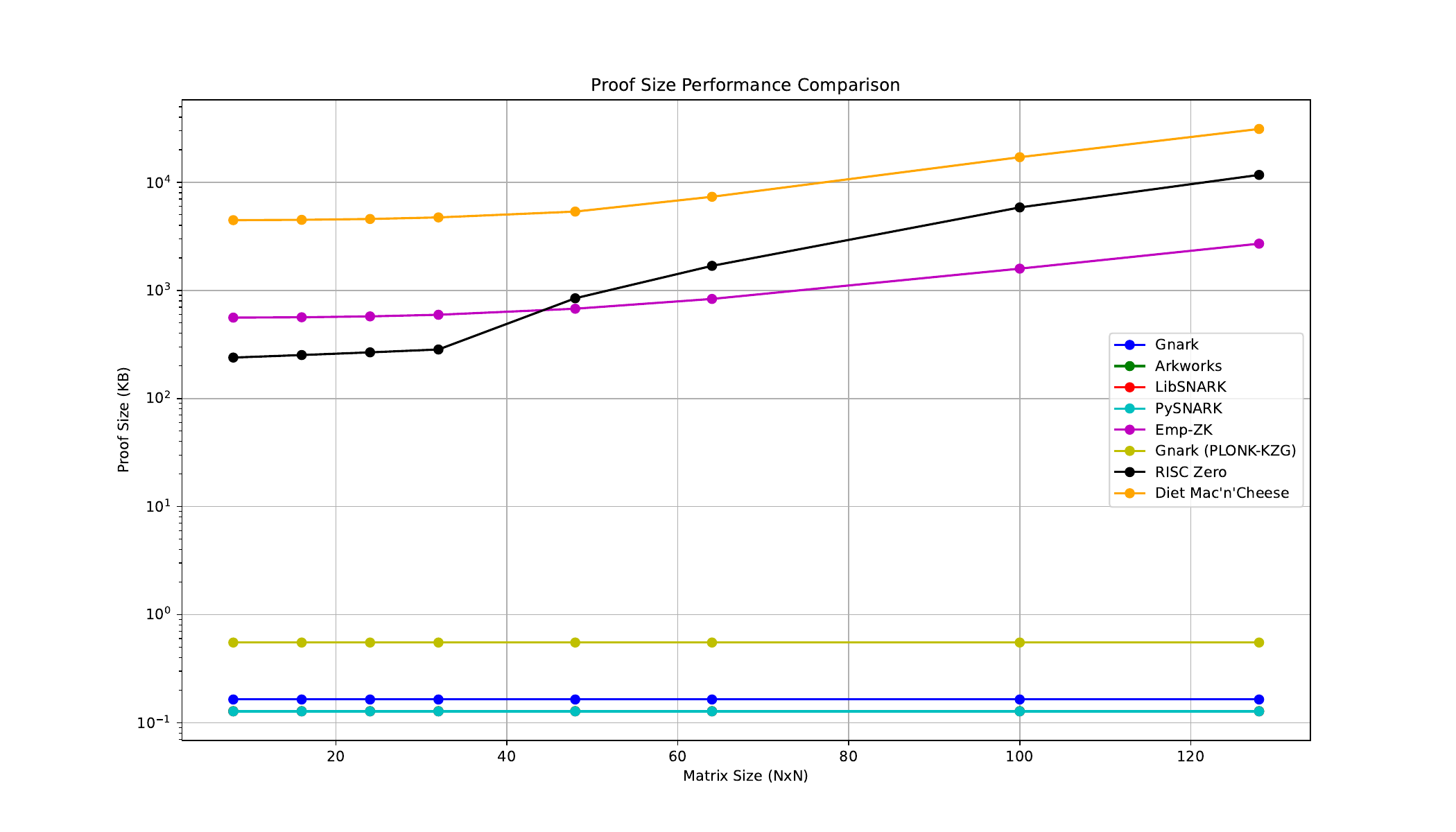}}
\hspace{1em}
\subfloat[Scaling of proof verification runtime.]{\includegraphics[width=.48\textwidth]{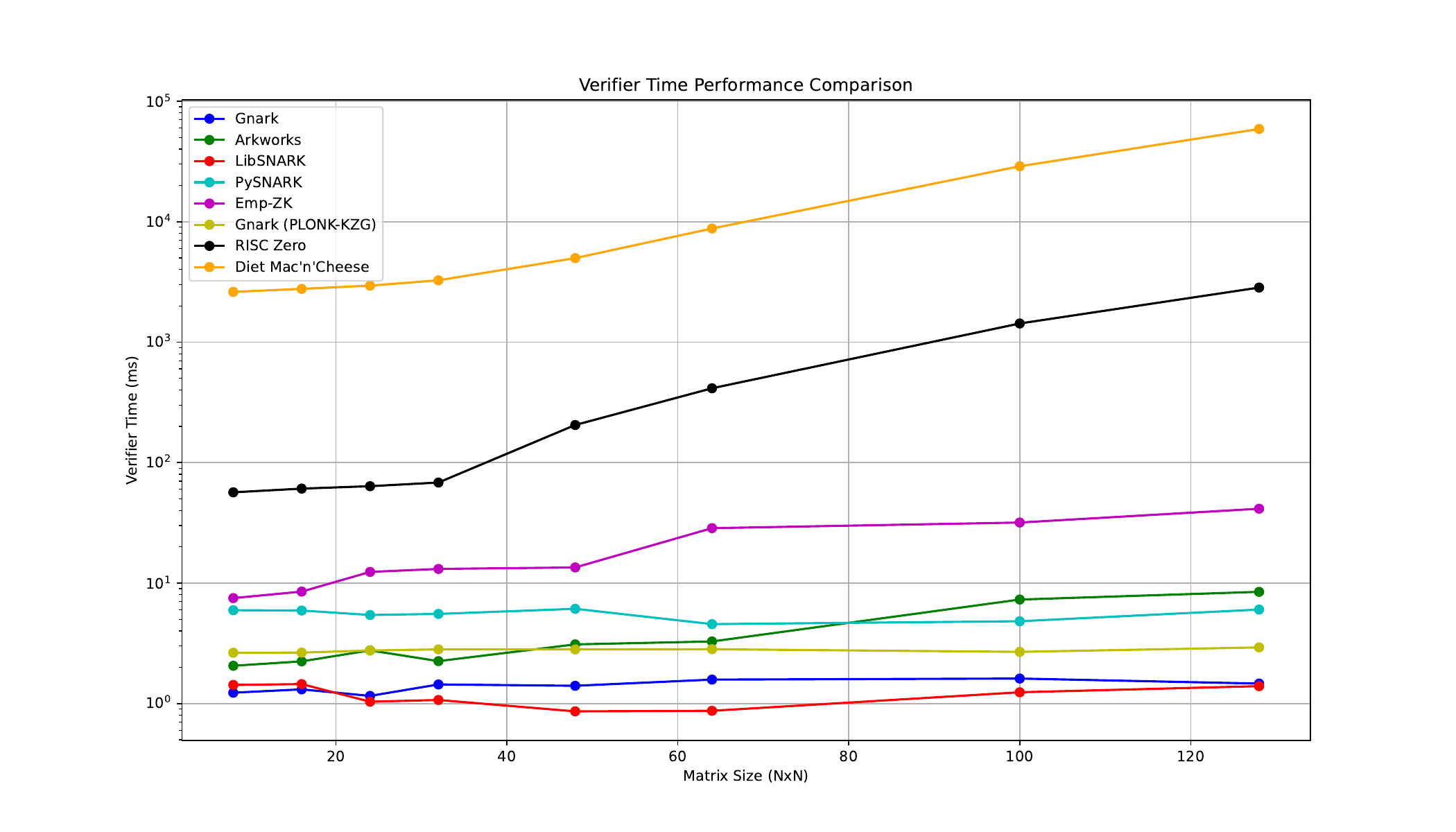}}
\caption{Analysis of scalability of select frameworks over matrix multiplication benchmarks spanning $8\times8$ matrix size, resulting in 64 constraints, up to $128\times128$ matrix size, resulting in 2,097,152 constraints.}
\label{fig:scalability}
\end{figure*}

In figure \ref{fig:scalability}, we perform an extensive scalability test on a subset of our evaluated frameworks to show how the trusted setup, proof generation and verification, and proof size/communication all scale as computation grows. We only do this for the matrix multiplication benchmark to avoid redundancies, and we believe it is sufficiently indicative of the framework's scalability. The subset of frameworks is chosen due to their promising performance, outlined in section \ref{sec:results}, and their ability to handle large circuits. Also,  some frameworks that were evaluated in section \ref{sec:results} ran into memory overflow issues on our machine as circuits scaled.

We found that proof size/communication and proof verification stay relatively constant for zk-SNARKs and \texttt{Gnark}'s PLONK implementation. Proof size and verification time follow very similar curves in the case of both \texttt{Emp-ZK} and \texttt{RISC Zero}. Most trusted setup times and proof generation times follow similar trajectories as circuits scale, except for \texttt{Emp-ZK}, which stays relatively constant. \texttt{Diet Mac'n'Cheese} provided us with interesting results that seem to separate it quite a bit from \texttt{Emp-ZK}, although they are both VOLE-based ZK solutions. Upon further inspection, we found that the only frontend that is available, \texttt{PicoZK}, serves as the main bottleneck. \texttt{PicoZK} compiles everything to the DARPA SIEVE Intermediate Representation (IR), which can be read by \texttt{Diet Mac'n'Cheese}. However, \texttt{Diet Mac'n'Cheese} is not optimized for operations presented in SIEVE IR format, which is why the presented scalability results make it seem like a less performant candidate. Upon our own code review and discussion with the \texttt{PicoZK} authors, we believe that further work into building a direct bridge between \texttt{PicoZK} and \texttt{Diet Mac'n'Cheese} can result in results that are much closer to \texttt{Emp-ZK}, while still being able to take advantage of \texttt{PicoZK}'s extremely user-friendly development process. \texttt{RISC Zero}'s proof generation time is several orders of magnitude larger than the other evaluated frameworks, however we believe this is due to it's nature as a zkVM, which requires extra underlying RISC-V-based operations to perform these tasks. Finally, we observe that zk-SNARKs exhibit essentially the same pattern of growth as circuits scale. \texttt{PySNARK} proof generation time grows drastically with circuit size, but we believe this is due to Python's compilation process, which results in slower runtimes. We highlight \texttt{Emp-ZK}, \texttt{Arkworks}, and \texttt{Gnark}'s zk-SNARK and PLONK implementation as excellent scalable frameworks for custom applications.

\textbf{Takeaways.} One of the main observations we make is the prevalance of usable zk-SNARK frameworks compared to all other constructions, and the lack of available tools for building systems with MPCitH ZKPs. This is primarily attributed to the fact that MPCitH ZKPs are primarily an academia-driven concept, meaning that the developers of the frameworks are not as concerned with commercialization, which naturally puts developer usability and accessibility to the wayside.
We also find that there is a lack of accessible dedicated PLONK-based Zk-SNARK frameworks. This is most likely due to the arithmetic and cryptographic flexibility that are attributed with PLONKs. Rather than building a dedicated PLONK framework, many developers, such as the developers of \texttt{SnarkJS} and \texttt{Gnark}, simply modify their frameworks to also support PLONK proving systems. As for zk-STARKs, we show that there are only a few accessible frameworks that can be utilized by developers. We do want to note that these frameworks are excellent, with a majority providing extensive documentation and very accessible frontends and APIs for those experienced with applied cryptography.

In our evaluation, we aim to keep settings (e.g. ECC curve, bitwidths, etc.) consistent between experiments as much as the frameworks allow us.
Nevertheless, it is not possible or fair to compare all of these frameworks to each other and choose a \textit{single} best one, as each different type of ZKP (zk-SNARKs, MPCitH, etc.) is built with different cryptographic, interactivity, and trusted setup assumptions. Rather, we analyze the results and discuss what application settings would benefit from each type of ZKP, and what frameworks can be used to realize those applications:

\textbf{zk-SNARKs: } Building an application powered by zk-SNARKs requires a computationally strong prover and trusted third party (for trusted setup), due to the underlying elliptic curve cryptographic assumption. As trusted setup parameters must be recomputed for every new circuit, zk-SNARKs are ideal when the computation remains relatively static. The perk of zk-SNARKs is that the proof size is succinct and somewhat constant and proofs are publicly-verifiable. zk-SNARKS are ideal for settings that are bandwidth-constrained and/or have resource-constrained verifiers. For building custom zk-SNARK-based applications, we recommend \textbf{Arkworks} or \textbf{Gnark}. These works provide excellent documentation, active development communities and forums, and many examples that can guide developers, while still maintaining competitive runtime and succinct proofs.

PLONKs have also proven to be a valuable tool for cryptographers, serving as the basis for some interesting applications \cite{0xPolygonZero2023Plonky2}. Due to the cryptographic flexibility, PLONKs are ideal for applications where available bandwidth and computation may require switching underlying cryptography. Alongside this, applications that can benefit from a single, versatile trusted setup are ideal candidates for PLONK-based design. For building custom PLONK-based applications, we recommend \textbf{GNARK-KZG}. The implementations provided in \texttt{Gnark}'s FRI and \texttt{SnarkJS} are valuable, however, as can be seen by the benchmarked results, they are not the most efficient. \texttt{Noir} offers detailed documentation and many examples, coupled with an active development community that are constantly presenting new applications built with Noir \cite{NoirLang2023AwesomeNoirBenchmarks}, however is outperformed by \texttt{GNARK-KZG}. We do note that \texttt{Noir} is also a fantastic option for developers. However, \texttt{GNARK-KZG} achieves excellent performance with an easily accessible API. 

\textbf{VOLE-based ZK: } Options for building VOLE-based ZK systems are admittedly limited, and the evaluations do not paint the full picture, as the presented runtimes are the amount of time that the prover and verifier must stay online, rather than time spent actively computing. VOLE-based ZK is an excellent candidates for custom applications when the setting naturally requires communication, such as distributed learning or the IoT, especially because they do not require trusted setup. These protocols also distribute the computational load between the prover and the verifier, rather than putting all of the work on the prover. For building custom VOLE-based ZK applications, we recommend \textbf{Emp-ZK}. This work provides great examples and readable C++ code, but most importantly, it supports arithmetic, Boolean, and mixed computation. Most importantly, due to the inclusion of Boolean circuit evaluation, this framework also supports floating point operations, which is not done by any other frameworks that we discuss. Although \texttt{Diet Mac'n'Cheese}, paired with \texttt{PicoZK}, is also a great solution, it still requires a bit more development before it is ready for application development without too many limitations (e.g. floating point support). However, we do note that this pairing can be used with quite an efficient development process, as it solely relies on readable Python code. Although it does present rather slow prover and verification times, we would like to emphasize that this is due to \texttt{PicoZK}'s compilation process being reliant on the SIEVE intermediate representation (IR) \cite{sieve}. \texttt{Diet Mac'n'Cheese} can operate over SIEVE IR files, but it has not been completely optimized for this process yet. While this is not a direct reflection of \texttt{Diet Mac'n'Cheese} performance, we believe it is currently the only solution to properly build ZKPs using the underlying \texttt{Diet Mac'n'Cheese} proof system. Nevertheless, the pairing of \texttt{Diet Mac'n'Cheese} with \texttt{PicoZK} provides the easiest route towards developing relatively complex applications that we have encountered through writing this paper.

\textbf{MPCitH.}
Once again, the options for building MPCitH ZKPs are quite limited. MPCitH ZKPs have most prominently been used in digital signatures \cite{PQC_MIRATH_Website, aragon2023ryde, SDITH_Website},, however, there are not many general-purpose frameworks available. For the purpose of this work, \textbf{Limbo} is the most accessible general-purpose framework for developing MPCitH ZK-based applications. This requires a baseline knowledge of Bristol fashion circuits \cite{bristol}, as this is how any computation \Cir is described in Limbo. However, once this hurdle is overcome, \texttt{Limbo} is a straightforward framework for building efficient MPCitH ZKPs. The main shortcoming of \texttt{Limbo} is its lack of support for anything but Boolean computation, which is why it performed poorly on the matrix multiplication benchmark. 



\textbf{zk-STARKs: } zk-STARKs have the potential to be an excellent solution for applications aiming to integrate ZKPs, however not many academic works have started integrating them into their applications, due to their relative nascency. Due to their lack of trusted setup and post-quantum, lightweight cryptography, they serve as great candidates for applications with strong provers and enough bandwidth to support proof transmission.
For building custom zk-STARK-based applications, we recommend \textbf{RISC Zero}. 
\texttt{Zilch} is another fantastic framework, however caused memory overflow for both of our benchmarks, which is why we do not consider it here. 
\texttt{RISC Zero} is the main offering from a startup, meaning the documentation is extensive and the framework is very well-developed. The main advantage of this framework is its ability to integrate with standard Rust crate easily, while still maintaining acceptable performance metrics. While the performance is a bit worse than \texttt{Miden VM}, it also do not require as much memory as \texttt{Miden VM} does, which caused memory overflow on a powerful server.

\begin{figure}
    \centering
    {\includegraphics[width=\columnwidth]{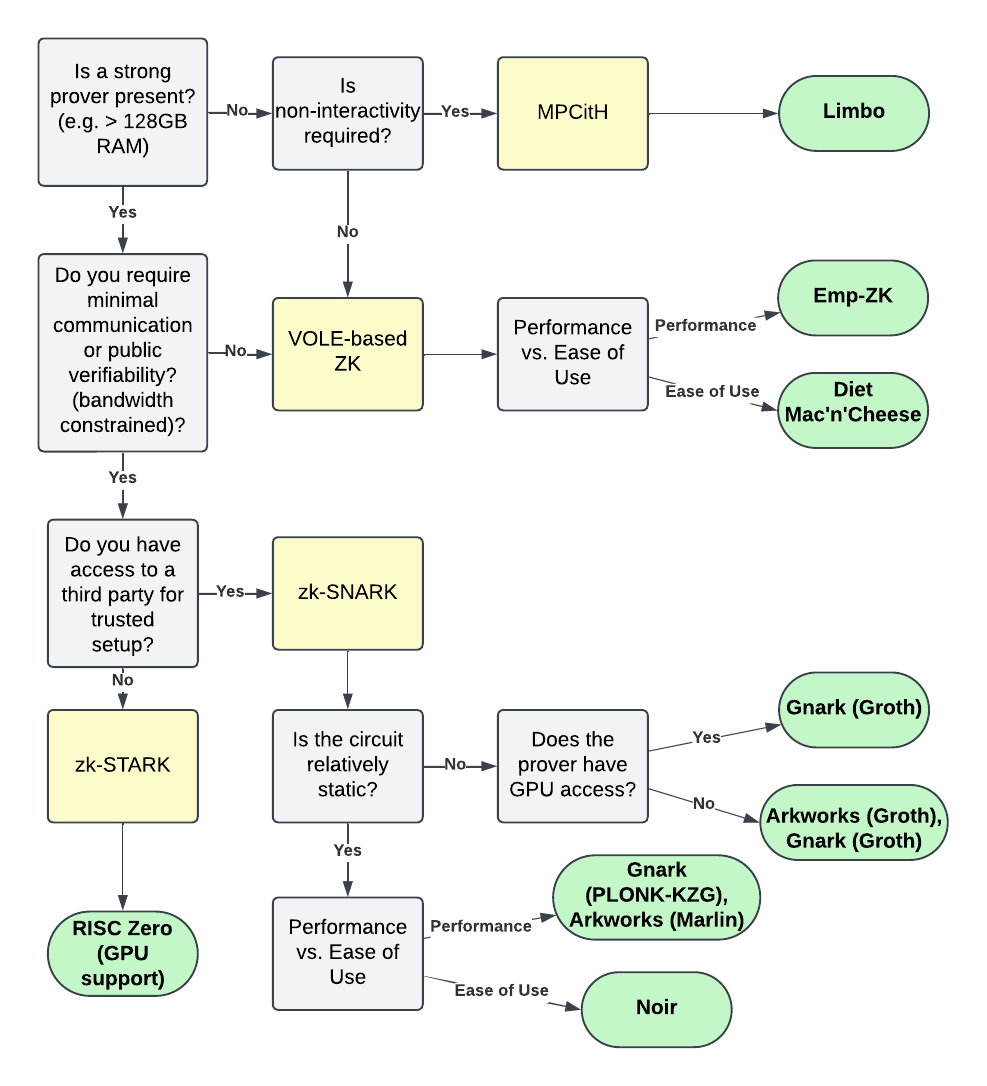}}
    \caption{Flow chart to guide users to the framework that best fits the requirements of their application and available resources.}
    \label{fig:flow}
\end{figure}

To accompany our suggestions, and to also take resource availability into account, figure \ref{fig:flow} provides a flow chart that developers can use to find the perfect framework for their applications, based on available resources and preferences. We note that most frameworks that we discuss in this survey can be used as solutions for building ZKP applications, however we believe that the highlighted frameworks in figure \ref{fig:flow} are the most promising and reliable.

\section{Discussion}

ZKPs in their current state have been used in cutting-edge applications, but there still remains a long path towards practicality. We highlight the novel ZKP applications in-depth in Appendix \ref{sec:applications}. The current challenges that hinder practical ZKP-based applications are three-fold:

\textbf{Usability.} As we've shown in this work, there are many, many great frameworks that enable the development of state-of-the-art ZKP, however the usability of many of them are hindered by their lack of a higher-level API or compatibility with a circuit description frontend, like Zokrates, Circom, or xJsnark. This makes the process of developing complex ZK-based applications much more difficult, as a user must learn the intricacies of a new protocol and the necessary syntax to take advantage of the promised performance.

\textbf{Accessibility.} Accessibility poses an issue, especially when evaluating academic frameworks. Academic frameworks often present state-of-the-art results, but normally lack documentation, examples, and other information that would allow for a user to replicate their results. This is not beneficial to the developers of the frameworks, as it presents a huge hurdle towards realizing the practicality of their proposed protocols in real-world applications. Frameworks that stem from industry are often, but not always, better in this respect, as the work is developed with a consumer in mind. The usability and accessibility issue can be solved by simply providing extensive documentation and attempting solutions that promote interoperability between similar frameworks. Another achievable solution is promoting communities where developers can discuss applications and solutions (e.g. Gitter). We acknowledge that there have been attempts to achieve interoperability for ZKP frameworks, such as CirC \cite{ozdemir2022circ} and zkinterface \cite{benarroch2019zkinterface}, however there still remains a lot of work to be done. 

\textbf{Performance.} Perhaps the most difficult problem with ZKPs to address is the computational overhead. The results we present in this work represent only tidbits of computation that would make up a full end-to-end ZK-based application. As can be seen, these state-of-the-art implementations still introduce a relatively large amount of overhead, even for simple tasks. This overhead only gets more overwhelming as applications get more complex. One realistic solution towards improving ZKP performance and reduce the computational overhead is hardware acceleration of state-of-the-art algorithms. This is a similar approach that is taken by the Intel HERACLES \cite{cammarota2022intel}, which is an attempt to build an end-to-end accelerator for FHE, with the goal of bringing FHE computation times down to the same order of magnitude as plaintext computation. One of the goals of this work is to provide a high-level overview of the state-of-the-art ZKP implementations, which is the first step toward identifying potential candidate protocols for acceleration. In addition to hardware acceleration, the continued exploration of algorithmic refinements and optimizations, such as the emerging research topic of GPU-based cryptographic optimizations \cite{lu2022cuzk, ma2023gzkp}, could lead to more efficient performance. There have also been solutions to propose custom hardware for end-to-end ZKP applications \cite{ahmed2024amaze, sheybani2025gotta} and proof generation \cite{samardzic2024accelerating, daftardar2024szkp}, however, they have not yet reached a stage of practicality.

The works highlighted in this paper represent the progression of algorithmic optimizations that have been instilled to bridge the gap between theory and practicality in the real-world application of ZKPs. While there are still significant hurdles that need to be overcome, many of them have actionable tasks, such as open-source framework developers providing clear examples and documentation, alongside an accessible API or general-purpose frontend compatibility. The future of ZKP lies in a collaborative effort across academia, industry, and the open-source community to address these challenges, leading to a landscape where ZKPs are not only theoretically profound but also a practically viable solution towards securing data and computations.

Our contributions toward this effort are this survey, alongside our provided open-source collection of development environments and accompanying examples for each framework, which will be actively maintained after publication of this paper. Our hope for this repository is to allow developers of new and existing frameworks to contribute Docker containers and examples programs upon release of their work. This provides a centralized hub for developers to test their custom applications on several different frameworks before making a final choice. The objective of this survey and accompanying repository is to demystify the ZK landscape for developers and new cryptographers and to significantly lower the barrier of entry to the field of ZKPs, while providing valuable insights as to which frameworks and constructions best suit their custom applications.


\section*{Acknowledgments}
This work was supported by DARPA Proofs under grant number HR0011-23-1-0006.

\section*{Conflict of Interests}
The authors declare that there is no conflict of interests regarding the publication of this paper.

{\footnotesize
\bibliographystyle{abbrv}
\bibliography{refs}

\begin{thebibliography}{100}

\bibitem{dietmc}
{Diet Mac'n'Cheese}.
\newblock \url{https://github.com/GaloisInc/swanky/tree/dev/diet-mac-and-cheese}.

\bibitem{hyraxZK}
{hyraxZK}.
\newblock \url{https://github.com/hyraxZK/hyraxZK}.

\bibitem{legosnark}
{LegoSNARK}.
\newblock \url{https://github.com/imdea-software/legosnark/}.

\bibitem{libsnark}
{libsnark}.
\newblock \url{https://github.com/scipr-lab/libsnark}.

\bibitem{Mirage}
{Mirage}.
\newblock \url{https://github.com/akosba/mirage/tree/master}.

\bibitem{picozk}
{PicoZK}.
\newblock \url{https://github.com/uvm-plaid/picozk}.

\bibitem{PySNARK}
{PySNARK}.
\newblock \url{https://github.com/meilof/pysnark}.

\bibitem{RapidSNARK}
{RapidSNARK}.
\newblock \url{https://github.com/iden3/rapidsnark}.

\bibitem{sieve}
{SIEVE Intermediate Representation}.
\newblock \url{https://github.com/sieve-zk/ir}.

\bibitem{Spartan}
{Spartan}.
\newblock \url{https://github.com/microsoft/Spartan}.

\bibitem{ristretto}
{The Ristretto Group}.
\newblock \url{https://ristretto.group}.

\bibitem{LambdaClass2023ArithmetizationSchemes}
Arithmetization schemes for zk-snarks, Jan. 2023.

\bibitem{DuskPlonk2023Rust}
dusk-plonk - rust, 2023.

\bibitem{Halo22023Book}
The halo2 book, 2023.

\bibitem{Halo2Club2023}
halo2.club, 2023.

\bibitem{RustImageCrate2023}
image, 2023.

\bibitem{Noir2023Documentation}
Introducing noir.
\newblock \url{https://noir-lang.org}, 2023.

\bibitem{PolygonMiden2023VMOverview}
Polygon miden vm overview, 2023.

\bibitem{Chainlink2023ZeroKnowledgeProof}
Zero-knowledge proof: Applications and use cases.
\newblock \url{https://chain.link/education-hub/zero-knowledge-proof-use-cases}, 2023.

\bibitem{PolygonMiden2023MidenVM}
0xPolygonMiden.
\newblock miden-vm, 2023.

\bibitem{0xPolygonZero2023Plonky2}
0xPolygonZero.
\newblock plonky2, 2023.

\bibitem{ahmed2024amaze}
A.~Ahmed, N.~Sheybani, D.~Moreno, N.~B. Njungle, T.~Gong, M.~Kinsy, and F.~Koushanfar.
\newblock Amaze: Accelerated mimc hardware architecture for zero-knowledge applications on the edge.
\newblock {\em arXiv preprint arXiv:2411.06350}, 2024.

\bibitem{mimchash}
M.~Albrecht, L.~Grassi, C.~Rechberger, A.~Roy, and T.~Tiessen.
\newblock Mimc: Efficient encryption and cryptographic hashing with minimal multiplicative complexity.
\newblock Cryptology ePrint Archive, Paper 2016/492, 2016.
\newblock \url{https://eprint.iacr.org/2016/492}.

\bibitem{ames2017ligero}
S.~Ames, C.~Hazay, Y.~Ishai, and M.~Venkitasubramaniam.
\newblock Ligero: Lightweight sublinear arguments without a trusted setup.
\newblock In {\em Proceedings of the 2017 acm sigsac conference on computer and communications security}, pages 2087--2104, 2017.

\bibitem{aragon2023ryde}
N.~Aragon, M.~Bardet, L.~Bidoux, J.-J. Chi-Dom{\'\i}nguez, V.~Dyseryn, T.~Feneuil, P.~Gaborit, A.~Joux, M.~Rivain, J.-P. Tillich, et~al.
\newblock Ryde specifications.
\newblock 2023.

\bibitem{arkworks}
arkworks contributors.
\newblock \texttt{arkworks} zksnark ecosystem, 2022.

\bibitem{ashur2018marvellous}
T.~Ashur and S.~Dhooghe.
\newblock Marvellous: a stark-friendly family of cryptographic primitives.
\newblock {\em Cryptology ePrint Archive}, 2018.

\bibitem{AztecProtocol2023Barretenberg}
AztecProtocol.
\newblock barretenberg, 2023.

\bibitem{babai1993transparent}
L.~Babai.
\newblock Transparent (holographic) proofs.
\newblock In {\em Annual Symposium on Theoretical Aspects of Computer Science}, pages 525--534. Springer, 1993.

\bibitem{bamberger2022verification}
K.~A. Bamberger, R.~Canetti, S.~Goldwasser, R.~Wexler, and E.~J. Zimmerman.
\newblock Verification dilemmas in law and the promise of zero-knowledge proofs.
\newblock {\em Berkeley Tech. LJ}, 37:1, 2022.

\bibitem{baum2023publicly}
C.~Baum, L.~Braun, C.~D. de~Saint~Guilhem, M.~Kloo{\ss}, E.~Orsini, L.~Roy, and P.~Scholl.
\newblock Publicly verifiable zero-knowledge and post-quantum signatures from vole-in-the-head.
\newblock In {\em Annual International Cryptology Conference}, pages 581--615. Springer, 2023.

\bibitem{baum2022moz}
C.~Baum, L.~Braun, A.~Munch-Hansen, and P.~Scholl.
\newblock Moz z 2 k arella: efficient vector-ole and zero-knowledge proofs over z 2 k.
\newblock In {\em Annual International Cryptology Conference}, pages 329--358. Springer, 2022.

\bibitem{baum2021mac}
C.~Baum, A.~J. Malozemoff, M.~B. Rosen, and P.~Scholl.
\newblock Mac'n'cheese: Zero-knowledge proofs for boolean and arithmetic circuits with nested disjunctions.
\newblock In {\em Advances in Cryptology--CRYPTO 2021: 41st Annual International Cryptology Conference, CRYPTO 2021, Virtual Event, August 16--20, 2021, Proceedings, Part IV 41}, pages 92--122. Springer, 2021.

\bibitem{baylina2020iden3}
J.~Baylina.
\newblock iden3/snarkjs, 2020.

\bibitem{pedersenhash}
J.~Baylina1 and M.~Belle`s.
\newblock 4-bit window pedersen hash on the baby jubjub elliptic curve.

\bibitem{belles2022circom}
M.~Bell{\'e}s-Mu{\~n}oz, M.~Isabel, J.~L. Mu{\~n}oz-Tapia, A.~Rubio, and J.~Baylina.
\newblock Circom: A circuit description language for building zero-knowledge applications.
\newblock {\em IEEE Transactions on Dependable and Secure Computing}, 2022.

\bibitem{ben2018fast}
E.~Ben-Sasson, I.~Bentov, Y.~Horesh, and M.~Riabzev.
\newblock Fast reed-solomon interactive oracle proofs of proximity.
\newblock In {\em 45th international colloquium on automata, languages, and programming (icalp 2018)}. Schloss Dagstuhl-Leibniz-Zentrum fuer Informatik, 2018.

\bibitem{ben2018scalable}
E.~Ben-Sasson, I.~Bentov, Y.~Horesh, and M.~Riabzev.
\newblock Scalable, transparent, and post-quantum secure computational integrity.
\newblock {\em Cryptology ePrint Archive}, 2018.

\bibitem{aurora}
E.~Ben-Sasson, A.~Chiesa, M.~Riabzev, N.~Spooner, M.~Virza, and N.~P. Ward.
\newblock Aurora: Transparent succinct arguments for r1cs.
\newblock Cryptology ePrint Archive, Paper 2018/828, 2018.
\newblock \url{https://eprint.iacr.org/2018/828}.

\bibitem{ben2014succinct}
E.~Ben-Sasson, A.~Chiesa, E.~Tromer, and M.~Virza.
\newblock Succinct $\{$Non-Interactive$\}$ zero knowledge for a von neumann architecture.
\newblock In {\em 23rd USENIX Security Symposium (USENIX Security 14)}, pages 781--796, 2014.

\bibitem{benadjila2024mq}
R.~Benadjila, T.~Feneuil, and M.~Rivain.
\newblock Mq on my mind: Post-quantum signatures from the non-structured multivariate quadratic problem.
\newblock In {\em 2024 IEEE 9th European Symposium on Security and Privacy (EuroS\&P)}, pages 468--485. IEEE, 2024.

\bibitem{benarroch2019zkinterface}
D.~Benarroch, K.~Gurkan, R.~Kahat, A.~Nicolas, and E.~Tromer.
\newblock zkinterface, a standard tool for zero-knowledge interoperability.
\newblock In {\em 2nd ZKProof Workshop. https://docs. zkproof. org/pages/standards/acceptedworkshop2/proposal--zk-interop-zkinterface. pdf}, 2019.

\bibitem{bernstein2006curve25519}
D.~J. Bernstein.
\newblock Curve25519: new diffie-hellman speed records.
\newblock In {\em Public Key Cryptography-PKC 2006: 9th International Conference on Theory and Practice in Public-Key Cryptography, New York, NY, USA, April 24-26, 2006. Proceedings 9}, pages 207--228. Springer, 2006.

\bibitem{bettaieb2024perk}
S.~Bettaieb, L.~Bidoux, V.~Dyseryn, A.~Esser, P.~Gaborit, M.~Kulkarni, and M.~Palumbi.
\newblock Perk: compact signature scheme based on a new variant of the permuted kernel problem.
\newblock {\em Designs, Codes and Cryptography}, pages 1--27, 2024.

\bibitem{binanceWhatZeroknowledge}
Binance.
\newblock {W}hat {I}s {Z}ero-knowledge {P}roof and {H}ow {D}oes {I}t {I}mpact {B}lockchain? | {B}inance {A}cademy --- academy.binance.com.
\newblock \url{https://academy.binance.com/en/articles/what-is-zero-knowledge-proof-and-how-does-it-impact-blockchain}.

\bibitem{bitansky2013recursive}
N.~Bitansky, R.~Canetti, A.~Chiesa, and E.~Tromer.
\newblock Recursive composition and bootstrapping for snarks and proof-carrying data.
\newblock In {\em Proceedings of the forty-fifth annual ACM symposium on Theory of computing}, pages 111--120, 2013.

\bibitem{bonawitz2017practical}
K.~Bonawitz, V.~Ivanov, B.~Kreuter, A.~Marcedone, H.~B. McMahan, S.~Patel, D.~Ramage, A.~Segal, and K.~Seth.
\newblock Practical secure aggregation for privacy-preserving machine learning.
\newblock In {\em proceedings of the 2017 ACM SIGSAC Conference on Computer and Communications Security}, pages 1175--1191, 2017.

\bibitem{boo2021litezkp}
E.~Boo, J.~Kim, and J.~Ko.
\newblock Litezkp: Lightening zero-knowledge proof-based blockchains for iot and edge platforms.
\newblock {\em IEEE Systems Journal}, 16(1):112--123, 2021.

\bibitem{gnark-v0.9.0}
G.~Botrel, T.~Piellard, Y.~E. Housni, I.~Kubjas, and A.~Tabaie.
\newblock Consensys/gnark: v0.9.0, Feb. 2023.

\bibitem{bowe2019recursive}
S.~Bowe, J.~Grigg, and D.~Hopwood.
\newblock Recursive proof composition without a trusted setup.
\newblock {\em Cryptology ePrint Archive}, 2019.

\bibitem{breidenbach2021chainlink}
L.~Breidenbach, C.~Cachin, B.~Chan, A.~Coventry, S.~Ellis, A.~Juels, F.~Koushanfar, A.~Miller, B.~Magauran, D.~Moroz, et~al.
\newblock Chainlink 2.0: Next steps in the evolution of decentralized oracle networks.
\newblock 2021.

\bibitem{ButerinQuadraticArithmeticPrograms}
V.~Buterin.
\newblock Quadratic arithmetic programs: From zero to hero, 2023.

\bibitem{cammarota2022intel}
R.~Cammarota.
\newblock Intel heracles: Homomorphic encryption revolutionary accelerator with correctness for learning-oriented end-to-end solutions.
\newblock In {\em Proceedings of the 2022 on Cloud Computing Security Workshop}, pages 3--3, 2022.

\bibitem{campanelli2019legosnark}
M.~Campanelli, D.~Fiore, and A.~Querol.
\newblock Legosnark: Modular design and composition of succinct zero-knowledge proofs.
\newblock In {\em Proceedings of the 2019 ACM SIGSAC Conference on Computer and Communications Security}, pages 2075--2092, 2019.

\bibitem{vcapko2022state}
D.~{\v{C}}apko, S.~Vukmirovi{\'c}, and N.~Nedi{\'c}.
\newblock State of the art of zero-knowledge proofs in blockchain.
\newblock In {\em 2022 30th Telecommunications Forum (TELFOR)}, pages 1--4. IEEE, 2022.

\bibitem{CelerNetwork2023Pantheon}
{Celer Network}.
\newblock The pantheon of zero knowledge proof development frameworks (updated!), 2023.

\bibitem{chainOverviewZeroKnowledge}
Chainlink.
\newblock {O}verview {O}f {Z}ero-{K}nowledge {B}lockchain {P}rojects | {C}hainlink --- chain.link.
\newblock \url{https://chain.link/education-hub/zero-knowledge-proof-projects}.

\bibitem{chiesa2020marlin}
A.~Chiesa, Y.~Hu, M.~Maller, P.~Mishra, N.~Vesely, and N.~Ward.
\newblock Marlin: Preprocessing zksnarks with universal and updatable srs.
\newblock In {\em Advances in Cryptology--EUROCRYPT 2020: 39th Annual International Conference on the Theory and Applications of Cryptographic Techniques, Zagreb, Croatia, May 10--14, 2020, Proceedings, Part I 39}, pages 738--768. Springer, 2020.

\bibitem{fractal}
A.~Chiesa, D.~Ojha, and N.~Spooner.
\newblock Fractal: Post-quantum and transparent recursive proofs from holography.
\newblock Cryptology ePrint Archive, Paper 2019/1076, 2019.
\newblock \url{https://eprint.iacr.org/2019/1076}.

\bibitem{ConsenSys2023Gnark}
{ConsenSys, Inc.}
\newblock gnark.
\newblock \url{https://docs.gnark.consensys.net/overview#gnark-is-fast}, 2023.

\bibitem{daftardar2024szkp}
A.~Daftardar, B.~Reagen, and S.~Garg.
\newblock Szkp: A scalable accelerator architecture for zero-knowledge proofs.
\newblock In {\em Proceedings of the 2024 International Conference on Parallel Architectures and Compilation Techniques}, pages 271--283, 2024.

\bibitem{uscs}
G.~Danezis, C.~Fournet, J.~Groth, and M.~Kohlweiss.
\newblock Square span programs with applications to succinct nizk arguments.
\newblock Cryptology ePrint Archive, Paper 2014/718, 2014.
\newblock \url{https://eprint.iacr.org/2014/718}.

\bibitem{limbo}
C.~D. de~Saint~Guilhem, E.~Orsini, and T.~Tanguy.
\newblock Limbo: Efficient zero-knowledge mpcith-based arguments.
\newblock Cryptology ePrint Archive, Paper 2021/215, 2021.
\newblock \url{https://eprint.iacr.org/2021/215}.

\bibitem{eberhardt2018zokrates}
J.~Eberhardt and S.~Tai.
\newblock Zokrates-scalable privacy-preserving off-chain computations.
\newblock In {\em 2018 IEEE International Conference on Internet of Things (iThings) and IEEE Green Computing and Communications (GreenCom) and IEEE Cyber, Physical and Social Computing (CPSCom) and IEEE Smart Data (SmartData)}, pages 1084--1091. IEEE, 2018.

\bibitem{fang2020local}
M.~Fang, X.~Cao, J.~Jia, and N.~Gong.
\newblock Local model poisoning attacks to $\{$Byzantine-Robust$\}$ federated learning.
\newblock In {\em 29th USENIX security symposium (USENIX Security 20)}, pages 1605--1622, 2020.

\bibitem{feng2021zen}
B.~Feng, L.~Qin, Z.~Zhang, Y.~Ding, and S.~Chu.
\newblock Zen: An optimizing compiler for verifiable, zero-knowledge neural network inferences.
\newblock {\em Cryptology ePrint Archive}, 2021.

\bibitem{fuchsbauer2018subversion}
G.~Fuchsbauer.
\newblock Subversion-zero-knowledge snarks.
\newblock In {\em Public-Key Cryptography--PKC 2018: 21st IACR International Conference on Practice and Theory of Public-Key Cryptography, Rio de Janeiro, Brazil, March 25-29, 2018, Proceedings, Part I 21}, pages 315--347. Springer, 2018.

\bibitem{gaba2022zero}
G.~S. Gaba, M.~Hedabou, P.~Kumar, A.~Braeken, M.~Liyanage, and M.~Alazab.
\newblock Zero knowledge proofs based authenticated key agreement protocol for sustainable healthcare.
\newblock {\em Sustainable Cities and Society}, 80:103766, 2022.

\bibitem{plonk}
A.~Gabizon, Z.~J. Williamson, and O.~Ciobotaru.
\newblock Plonk: Permutations over lagrange-bases for oecumenical noninteractive arguments of knowledge.
\newblock Cryptology ePrint Archive, Paper 2019/953, 2019.
\newblock \url{https://eprint.iacr.org/2019/953}.

\bibitem{swanky}
{Galois, Inc.}
\newblock {swanky}: A suite of rust libraries for secure computation.
\newblock \url{https://github.com/GaloisInc/swanky}, 2019.

\bibitem{ganesh2023rinocchio}
C.~Ganesh, A.~Nitulescu, and E.~Soria-Vazquez.
\newblock Rinocchio: Snarks for ring arithmetic.
\newblock {\em Journal of Cryptology}, 36(4):41, 2023.

\bibitem{gennaro2013quadratic}
R.~Gennaro, C.~Gentry, B.~Parno, and M.~Raykova.
\newblock Quadratic span programs and succinct nizks without pcps.
\newblock In {\em Advances in Cryptology--EUROCRYPT 2013: 32nd Annual International Conference on the Theory and Applications of Cryptographic Techniques, Athens, Greece, May 26-30, 2013. Proceedings 32}, pages 626--645. Springer, 2013.

\bibitem{ghodsi2023zprobe}
Z.~Ghodsi, M.~Javaheripi, N.~Sheybani, X.~Zhang, K.~Huang, and F.~Koushanfar.
\newblock zprobe: Zero peek robustness checks for federated learning.
\newblock In {\em Proceedings of the IEEE/CVF International Conference on Computer Vision}, pages 4860--4870, 2023.

\bibitem{goldreich1994definitions}
O.~Goldreich and Y.~Oren.
\newblock Definitions and properties of zero-knowledge proof systems.
\newblock {\em Journal of Cryptology}, 7(1):1--32, 1994.

\bibitem{goldwasser2015delegating}
S.~Goldwasser, Y.~T. Kalai, and G.~N. Rothblum.
\newblock Delegating computation: interactive proofs for muggles.
\newblock {\em Journal of the ACM (JACM)}, 62(4):1--64, 2015.

\bibitem{goldwasser2019knowledge}
S.~Goldwasser, S.~Micali, and C.~Rackoff.
\newblock The knowledge complexity of interactive proof-systems.
\newblock In {\em Providing sound foundations for cryptography: On the work of shafi goldwasser and silvio micali}, pages 203--225. 2019.

\bibitem{golovnev2023brakedown}
A.~Golovnev, J.~Lee, S.~Setty, J.~Thaler, and R.~S. Wahby.
\newblock Brakedown: Linear-time and field-agnostic snarks for r1cs.
\newblock In {\em Annual International Cryptology Conference}, pages 193--226. Springer, 2023.

\bibitem{poseidonhash}
L.~Grassi, D.~Khovratovich, C.~Rechberger, A.~Roy, and M.~Schofnegger.
\newblock Poseidon: A new hash function for {Zero-Knowledge} proof systems.
\newblock In {\em 30th USENIX Security Symposium (USENIX Security 21)}, pages 519--535. USENIX Association, Aug. 2021.

\bibitem{groth16}
J.~Groth.
\newblock On the size of pairing-based non-interactive arguments.
\newblock In M.~Fischlin and J.-S. Coron, editors, {\em Advances in Cryptology -- EUROCRYPT 2016}, pages 305--326, Berlin, Heidelberg, 2016. Springer Berlin Heidelberg.

\bibitem{cryptoeprint:2018/280}
J.~Groth, M.~Kohlweiss, M.~Maller, S.~Meiklejohn, and I.~Miers.
\newblock Updatable and universal common reference strings with applications to zk-snarks.
\newblock Cryptology ePrint Archive, Paper 2018/280, 2018.
\newblock \url{https://eprint.iacr.org/2018/280}.

\bibitem{groth2017snarky}
J.~Groth and M.~Maller.
\newblock Snarky signatures: Minimal signatures of knowledge from simulation-extractable snarks.
\newblock In {\em Annual International Cryptology Conference}, pages 581--612. Springer, 2017.

\bibitem{grubbs2022zero}
P.~Grubbs, A.~Arun, Y.~Zhang, J.~Bonneau, and M.~Walfish.
\newblock $\{$Zero-Knowledge$\}$ middleboxes.
\newblock In {\em 31st USENIX Security Symposium (USENIX Security 22)}, pages 4255--4272, 2022.

\bibitem{habock2022summary}
U.~Hab{\"o}ck.
\newblock A summary on the fri low degree test.
\newblock {\em Cryptology ePrint Archive}, 2022.

\bibitem{hastings2019sok}
M.~Hastings, B.~Hemenway, D.~Noble, and S.~Zdancewic.
\newblock Sok: General purpose compilers for secure multi-party computation.
\newblock In {\em 2019 IEEE symposium on security and privacy (SP)}, pages 1220--1237. IEEE, 2019.

\bibitem{hopwood2016zcash}
D.~Hopwood, S.~Bowe, T.~Hornby, N.~Wilcox, et~al.
\newblock Zcash protocol specification.
\newblock {\em GitHub: San Francisco, CA, USA}, 4(220):32, 2016.

\bibitem{icicle}
Ingonyama.
\newblock Icicle: Gpu library for zk acceleration.

\bibitem{ishai2007zero}
Y.~Ishai, E.~Kushilevitz, R.~Ostrovsky, and A.~Sahai.
\newblock Zero-knowledge from secure multiparty computation.
\newblock In {\em Proceedings of the thirty-ninth annual ACM symposium on Theory of computing}, pages 21--30, 2007.

\bibitem{juels2024props}
A.~Juels and F.~Koushanfar.
\newblock Props for machine-learning security.
\newblock {\em arXiv preprint arXiv:2410.20522}, 2024.

\bibitem{kate2010constant}
A.~Kate, G.~M. Zaverucha, and I.~Goldberg.
\newblock Constant-size commitments to polynomials and their applications.
\newblock In {\em Advances in Cryptology-ASIACRYPT 2010: 16th International Conference on the Theory and Application of Cryptology and Information Security, Singapore, December 5-9, 2010. Proceedings 16}, pages 177--194. Springer, 2010.

\bibitem{kilian1992note}
J.~Kilian.
\newblock A note on efficient zero-knowledge proofs and arguments.
\newblock In {\em Proceedings of the twenty-fourth annual ACM symposium on Theory of computing}, pages 723--732, 1992.

\bibitem{kosba2020mirage}
A.~Kosba, D.~Papadopoulos, C.~Papamanthou, and D.~Song.
\newblock $\{$MIRAGE$\}$: Succinct arguments for randomized algorithms with applications to universal $\{$zk-SNARKs$\}$.
\newblock In {\em 29th USENIX Security Symposium (USENIX Security 20)}, pages 2129--2146, 2020.

\bibitem{kosba2018xjsnark}
A.~Kosba, C.~Papamanthou, and E.~Shi.
\newblock xjsnark: A framework for efficient verifiable computation.
\newblock In {\em 2018 IEEE Symposium on Security and Privacy (SP)}, pages 944--961. IEEE, 2018.

\bibitem{kothapalli2022nova}
A.~Kothapalli, S.~Setty, and I.~Tzialla.
\newblock Nova: Recursive zero-knowledge arguments from folding schemes.
\newblock In {\em Annual International Cryptology Conference}, pages 359--388. Springer, 2022.

\bibitem{KULeuvenCOSIC2023Limbo}
KULeuven-COSIC.
\newblock Limbo, 2023.

\bibitem{landau1988zero}
S.~Landau.
\newblock Zero knowledge and the department of defense.
\newblock {\em Notices of the American Mathematical Society}, 35(1):5--12, 1988.

\bibitem{lee2020vcnn}
S.~Lee, H.~Ko, J.~Kim, and H.~Oh.
\newblock vcnn: Verifiable convolutional neural network based on zk-snarks.
\newblock {\em Cryptology ePrint Archive}, 2020.

\bibitem{lin2021efficient}
C.~Lin, M.~Luo, X.~Huang, K.-K.~R. Choo, and D.~He.
\newblock An efficient privacy-preserving credit score system based on noninteractive zero-knowledge proof.
\newblock {\em IEEE systems journal}, 16(1):1592--1601, 2021.

\bibitem{lipmaa2016prover}
H.~Lipmaa.
\newblock Prover-efficient commit-and-prove zero-knowledge snarks.
\newblock In {\em Progress in Cryptology--AFRICACRYPT 2016: 8th International Conference on Cryptology in Africa, Fes, Morocco, April 13-15, 2016, Proceedings 8}, pages 185--206. Springer, 2016.

\bibitem{liu2021zkcnn}
T.~Liu, X.~Xie, and Y.~Zhang.
\newblock Zkcnn: Zero knowledge proofs for convolutional neural network predictions and accuracy.
\newblock In {\em Proceedings of the 2021 ACM SIGSAC Conference on Computer and Communications Security}, pages 2968--2985, 2021.

\bibitem{lu2022cuzk}
T.~Lu, C.~Wei, R.~Yu, C.~Chen, W.~Fang, L.~Wang, Z.~Wang, and W.~Chen.
\newblock Cuzk: Accelerating zero-knowledge proof with a faster parallel multi-scalar multiplication algorithm on gpus.
\newblock {\em Cryptology ePrint Archive}, 2022.

\bibitem{lycklama2023rofl}
H.~Lycklama, L.~Burkhalter, A.~Viand, N.~K{\"u}chler, and A.~Hithnawi.
\newblock Rofl: Robustness of secure federated learning.
\newblock In {\em 2023 IEEE Symposium on Security and Privacy (SP)}, pages 453--476. IEEE, 2023.

\bibitem{ma2023gzkp}
W.~Ma, Q.~Xiong, X.~Shi, X.~Ma, H.~Jin, H.~Kuang, M.~Gao, Y.~Zhang, H.~Shen, and W.~Hu.
\newblock Gzkp: A gpu accelerated zero-knowledge proof system.
\newblock In {\em Proceedings of the 28th ACM International Conference on Architectural Support for Programming Languages and Operating Systems, Volume 2}, pages 340--353, 2023.

\bibitem{cryptoeprint:2019/099}
M.~Maller, S.~Bowe, M.~Kohlweiss, and S.~Meiklejohn.
\newblock Sonic: Zero-knowledge snarks from linear-size universal and updateable structured reference strings.
\newblock Cryptology ePrint Archive, Paper 2019/099, 2019.
\newblock \url{https://eprint.iacr.org/2019/099}.

\bibitem{Monero2023}
Monero.
\newblock Home | monero - secure, private, untraceable.
\newblock \url{https://www.getmonero.org}, 2023.

\bibitem{mouris2021zilch}
D.~Mouris and N.~G. Tsoutsos.
\newblock Zilch: A framework for deploying transparent zero-knowledge proofs.
\newblock {\em IEEE Transactions on Information Forensics and Security}, 16:3269--3284, 2021.

\bibitem{munoz2022circom}
J.~L. Mu{\~n}oz-Tapia, M.~Belles, M.~Isabel, A.~Rubio, and J.~Baylina.
\newblock Circom: A robust and scalable language for building complex zero-knowledge circuits.
\newblock 2022.

\bibitem{NIST_PQC_Round2_Signatures}
{National Institute of Standards and Technology (NIST)}.
\newblock Round 2 additional signatures.
\newblock \url{https://csrc.nist.gov/projects/pqc-dig-sig/round-2-additional-signatures}, 2020.

\bibitem{NoirLang2023AwesomeNoirBenchmarks}
Noir-Lang.
\newblock Benchmarks in awesome-noir, 2023.

\bibitem{ozdemir2022circ}
A.~Ozdemir, F.~Brown, and R.~S. Wahby.
\newblock Circ: Compiler infrastructure for proof systems, software verification, and more.
\newblock In {\em 2022 IEEE Symposium on Security and Privacy (SP)}, pages 2248--2266. IEEE, 2022.

\bibitem{parno2016pinocchio}
B.~Parno, J.~Howell, C.~Gentry, and M.~Raykova.
\newblock Pinocchio: Nearly practical verifiable computation.
\newblock {\em Communications of the ACM}, 59(2):103--112, 2016.

\bibitem{partala2020non}
J.~Partala, T.~H. Nguyen, and S.~Pirttikangas.
\newblock Non-interactive zero-knowledge for blockchain: A survey.
\newblock {\em IEEE Access}, 8:227945--227961, 2020.

\bibitem{PolygonLabs2023PolygonZkEVM}
{Polygon Labs UI (Cayman) Ltd.}
\newblock Polygon zkevm | scaling for the ethereum virtual machine.
\newblock \url{https://polygon.technology/polygon-zkevm}, 2023.

\bibitem{PQC_MIRATH_Website}
{PQC-MIRATH Consortium}.
\newblock Pqc-mirath.
\newblock \url{https://pqc-mirath.org/}, 2023.

\bibitem{rabin2012strictly}
M.~O. Rabin, Y.~Mansour, S.~Muthukrishnan, and M.~Yung.
\newblock Strictly-black-box zero-knowledge and efficient validation of financial transactions.
\newblock In {\em International Colloquium on Automata, Languages, and Programming}, pages 738--749. Springer, 2012.

\bibitem{RISCZero2023DeveloperDocs}
{RISC Zero, Inc.}
\newblock Introduction | risc zero developer docs, 2023.

\bibitem{roy2022eiffel}
A.~Roy~Chowdhury, C.~Guo, S.~Jha, and L.~van~der Maaten.
\newblock Eiffel: Ensuring integrity for federated learning.
\newblock In {\em Proceedings of the 2022 ACM SIGSAC Conference on Computer and Communications Security}, pages 2535--2549, 2022.

\bibitem{samardzic2024accelerating}
N.~Samardzic, S.~Langowski, S.~Devadas, and D.~Sanchez.
\newblock Accelerating zero-knowledge proofs through hardware-algorithm co-design.
\newblock In {\em 2024 57th IEEE/ACM International Symposium on Microarchitecture (MICRO)}, pages 366--379. IEEE, 2024.

\bibitem{SciprLab2023Libiop}
SciprLab.
\newblock libiop, 2023.

\bibitem{SDITH_Website}
{SDITH}.
\newblock Sdith.
\newblock \url{https://sdith.org/index.html}, 2023.

\bibitem{setty2020spartan}
S.~Setty.
\newblock Spartan: Efficient and general-purpose zksnarks without trusted setup.
\newblock In {\em Annual International Cryptology Conference}, pages 704--737. Springer, 2020.

\bibitem{sharma2020blockchain}
B.~Sharma, R.~Halder, and J.~Singh.
\newblock Blockchain-based interoperable healthcare using zero-knowledge proofs and proxy re-encryption.
\newblock In {\em 2020 International Conference on COMmunication Systems \& NETworkS (COMSNETS)}, pages 1--6. IEEE, 2020.

\bibitem{sheybani2023zkrownn}
N.~Sheybani, Z.~Ghodsi, R.~Kapila, and F.~Koushanfar.
\newblock Zkrownn: Zero knowledge right of ownership for neural networks.
\newblock In {\em 2023 60th ACM/IEEE Design Automation Conference (DAC)}, pages 1--6. IEEE, 2023.

\bibitem{sheybani2025gotta}
N.~Sheybani, T.~Gong, A.~Ahmed, N.~B. Njungle, M.~Kinsy, and F.~Koushanfar.
\newblock Gotta hash'em all! speeding up hash functions for zero-knowledge proof applications.
\newblock {\em arXiv preprint arXiv:2501.18780}, 2025.

\bibitem{sidorenco2021formal}
N.~Sidorenco, S.~Oechsner, and B.~Spitters.
\newblock Formal security analysis of mpc-in-the-head zero-knowledge protocols.
\newblock In {\em 2021 IEEE 34th Computer Security Foundations Symposium (CSF)}, pages 1--14. IEEE, 2021.

\bibitem{vsimunic2021verifiable}
S.~{\v{S}}imuni{\'c}, D.~Bernaca, and K.~Lenac.
\newblock Verifiable computing applications in blockchain.
\newblock {\em IEEE Access}, 9:156729--156745, 2021.

\bibitem{so2020byzantine}
J.~So, B.~G{\"u}ler, and A.~S. Avestimehr.
\newblock Byzantine-resilient secure federated learning.
\newblock {\em IEEE Journal on Selected Areas in Communications}, 39(7):2168--2181, 2020.

\bibitem{9520375}
X.~Sun, F.~R. Yu, P.~Zhang, Z.~Sun, W.~Xie, and X.~Peng.
\newblock A survey on zero-knowledge proof in blockchain.
\newblock {\em IEEE Network}, 35(4):198--205, 2021.

\bibitem{SunblazeUCB2023Virgo}
sunblaze ucb.
\newblock Virgo, 2023.

\bibitem{thorpe2009zero}
C.~Thorpe and D.~C. Parkes.
\newblock Zero-knowledge proofs in large trades, July~9 2009.
\newblock US Patent App. 12/261,249.

\bibitem{Delendum2023ZKSystemBenchmarking}
B.~Threadbare, D.~Schmid, T.~Carstens, B.~Retford, D.~Lubarov, A.~Nagornyi, and V.~Tan.
\newblock Zk system benchmarking, 2023.

\bibitem{bristol}
S.~Tillich and N.~Smart.
\newblock (bristol format) circuits of basic functions suitable for mpc and fhe, 2023.

\bibitem{tomaz2020preserving}
A.~E.~B. Tomaz, J.~C. Do~Nascimento, A.~S. Hafid, and J.~N. De~Souza.
\newblock Preserving privacy in mobile health systems using non-interactive zero-knowledge proof and blockchain.
\newblock {\em IEEE access}, 8:204441--204458, 2020.

\bibitem{TrustworthyComputing2023Zilch}
TrustworthyComputing.
\newblock Zilch, 2023.

\bibitem{viand2021sok}
A.~Viand, P.~Jattke, and A.~Hithnawi.
\newblock Sok: Fully homomorphic encryption compilers.
\newblock In {\em 2021 IEEE Symposium on Security and Privacy (SP)}, pages 1092--1108. IEEE, 2021.

\bibitem{viand2023verifiable}
A.~Viand, C.~Knabenhans, and A.~Hithnawi.
\newblock Verifiable fully homomorphic encryption.
\newblock {\em arXiv preprint arXiv:2301.07041}, 2023.

\bibitem{wahby2017full}
R.~S. Wahby, Y.~Ji, A.~J. Blumberg, A.~Shelat, J.~Thaler, M.~Walfish, and T.~Wies.
\newblock Full accounting for verifiable outsourcing.
\newblock In {\em Proceedings of the 2017 ACM SIGSAC Conference on Computer and Communications Security}, pages 2071--2086, 2017.

\bibitem{wahby2018doubly}
R.~S. Wahby, I.~Tzialla, A.~Shelat, J.~Thaler, and M.~Walfish.
\newblock Doubly-efficient zksnarks without trusted setup.
\newblock In {\em 2018 IEEE Symposium on Security and Privacy (SP)}, pages 926--943. IEEE, 2018.

\bibitem{emptool}
X.~Wang.
\newblock Emp-toolkit.

\bibitem{empzk}
X.~Wang.
\newblock Emp-zk.

\bibitem{Weng2023VOLEBasedInteractive}
C.~Weng, A.~Coventry, S.~Hussain, D.~Malkhi, A.~Topliceanu, X.~Wang, and F.~Zhang.
\newblock Vole-based interactive commitments, Jan. 2023.

\bibitem{weng2021wolverine}
C.~Weng, K.~Yang, J.~Katz, and X.~Wang.
\newblock Wolverine: fast, scalable, and communication-efficient zero-knowledge proofs for boolean and arithmetic circuits.
\newblock In {\em 2021 IEEE Symposium on Security and Privacy (SP)}, pages 1074--1091. IEEE, 2021.

\bibitem{weng2021mystique}
C.~Weng, K.~Yang, X.~Xie, J.~Katz, and X.~Wang.
\newblock Mystique: Efficient conversions for $\{$Zero-Knowledge$\}$ proofs with applications to machine learning.
\newblock In {\em 30th USENIX Security Symposium (USENIX Security 21)}, pages 501--518, 2021.

\bibitem{wu2014survey}
H.~Wu, F.~Wang, et~al.
\newblock A survey of noninteractive zero knowledge proof system and its applications.
\newblock {\em The Scientific World Journal}, 2014, 2014.

\bibitem{yang2021quicksilver}
K.~Yang, P.~Sarkar, C.~Weng, and X.~Wang.
\newblock Quicksilver: Efficient and affordable zero-knowledge proofs for circuits and polynomials over any field.
\newblock {\em IACR Cryptol. ePrint Arch.}, 2021:76, 2021.

\bibitem{zhang2020deco}
F.~Zhang, D.~Maram, H.~Malvai, S.~Goldfeder, and A.~Juels.
\newblock Deco: Liberating web data using decentralized oracles for tls.
\newblock In {\em Proceedings of the 2020 ACM SIGSAC Conference on Computer and Communications Security}, pages 1919--1938, 2020.

\bibitem{zhang2020zero}
J.~Zhang, Z.~Fang, Y.~Zhang, and D.~Song.
\newblock Zero knowledge proofs for decision tree predictions and accuracy.
\newblock In {\em Proceedings of the 2020 ACM SIGSAC Conference on Computer and Communications Security}, pages 2039--2053, 2020.

\bibitem{virgoplus}
J.~Zhang, T.~Liu, W.~Wang, Y.~Zhang, D.~Song, X.~Xie, and Y.~Zhang.
\newblock Doubly efficient interactive proofs for general arithmetic circuits with linear prover time.
\newblock Cryptology ePrint Archive, Paper 2020/1247, 2020.
\newblock \url{https://eprint.iacr.org/2020/1247}.

\bibitem{zhang2020transparent}
J.~Zhang, T.~Xie, Y.~Zhang, and D.~Song.
\newblock Transparent polynomial delegation and its applications to zero knowledge proof.
\newblock In {\em 2020 IEEE Symposium on Security and Privacy (SP)}, pages 859--876. IEEE, 2020.

\bibitem{ZkonduitInc2023EZKL}
{Zkonduit Inc.}
\newblock What is ezkl?, 2023.

\end{thebibliography}
}

\begin{IEEEbiography}[{\includegraphics[width=1in,height=1.25in,clip,keepaspectratio]{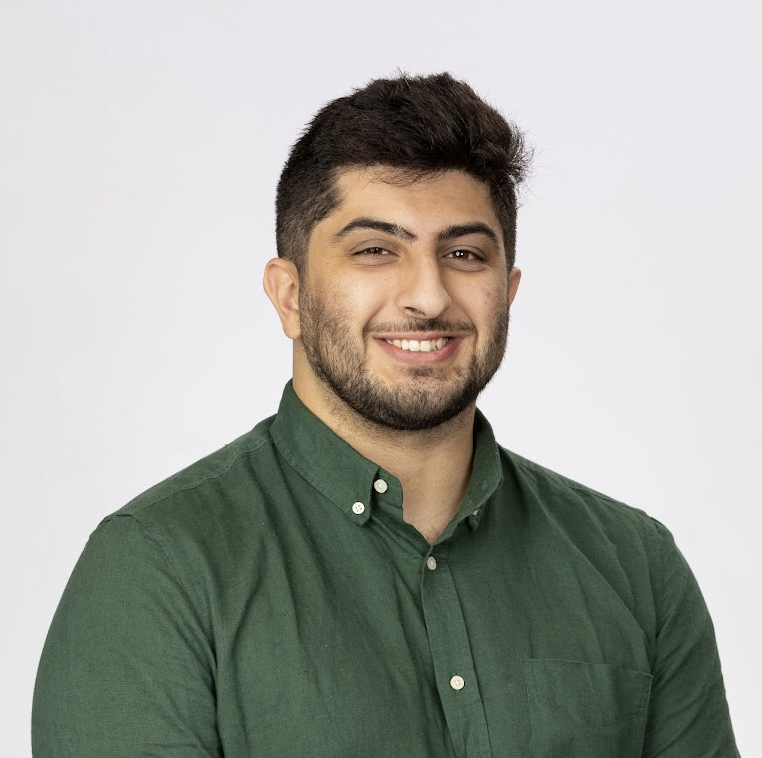}}]{Nojan Sheybani} is a Ph.D. candidate in the department of Electrical and Computer Engineering (ECE) at the University of California San Diego (UCSD). His research is focused on applied cryptography, hardware/software co-design, and zero-knowledge proofs. In particular, a common theme in his work is the application and optimization of privacy-preserving techniques to build practical and secure real-world systems.
\end{IEEEbiography}

\begin{IEEEbiography}[{\includegraphics[width=1in,height=1.25in,clip,keepaspectratio]{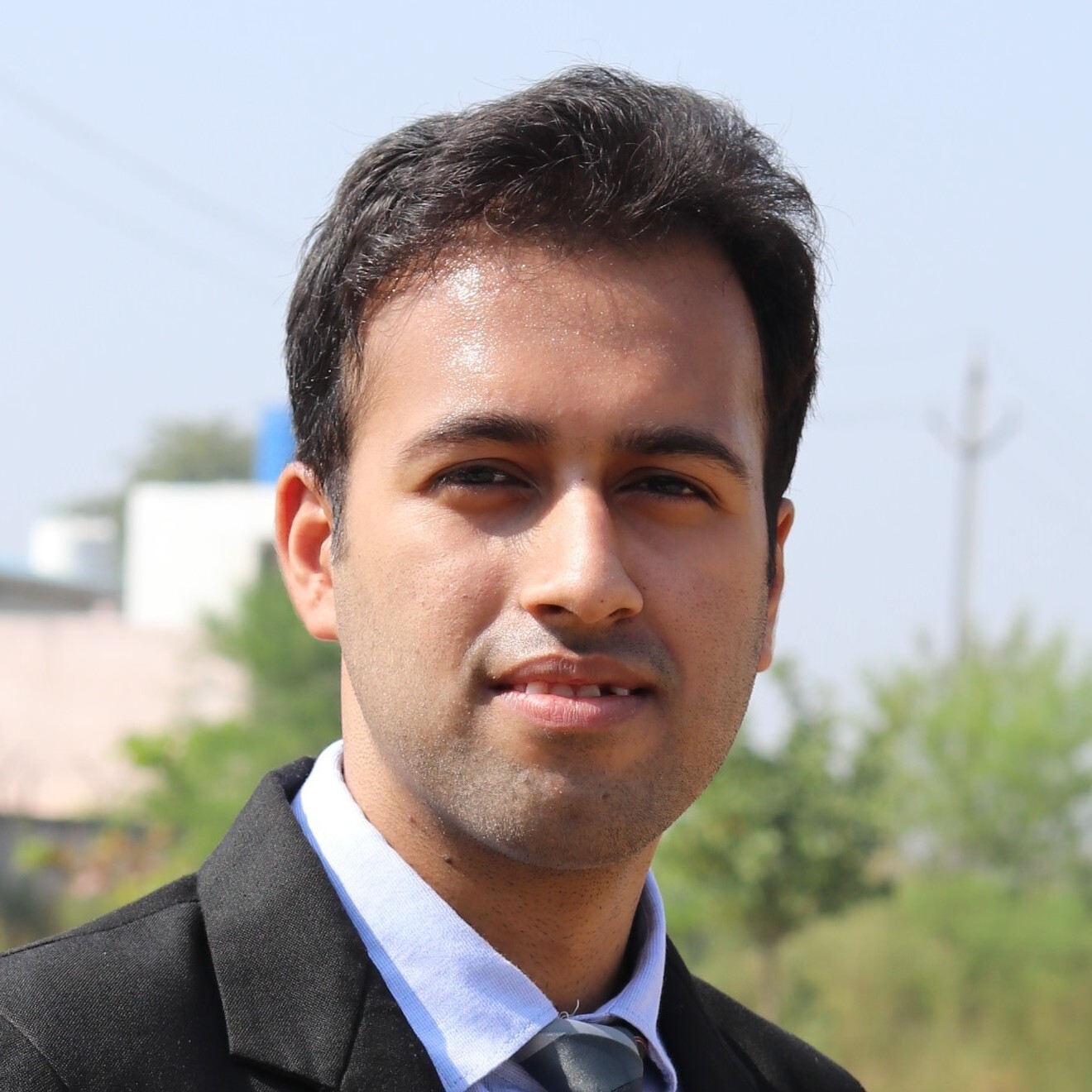}}]{Anees Ahmed} received his M.S. degree in Computer Science from Arizona State University. His research interests include privacy-preserving computation, zero-knowledge proofs, hardware acceleration, and computer architecture. Prior to his master's degree, he was a software engineer for two years. 
\end{IEEEbiography}

\begin{IEEEbiography}[{\includegraphics[width=1in,height=1.25in,clip,keepaspectratio]{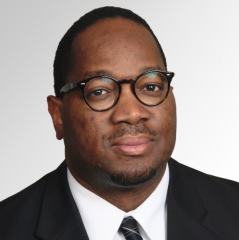}}]{Michel Kinsy} is an associate professor in the School of Computing and Augmented Intelligence and the director of the Secure, Trusted, and Assured Microelectronics Center. He focuses his research on microelectronics security, secure processors and systems design, hardware security, and efficient hardware design and implementation of post-quantum cryptography systems. Kinsy is an MIT Presidential Fellow and a CRA-WP Inaugural Skip Ellis Career Award recipient.
\end{IEEEbiography}

\begin{IEEEbiography}[{\includegraphics[width=1in,height=1.25in,clip,keepaspectratio]{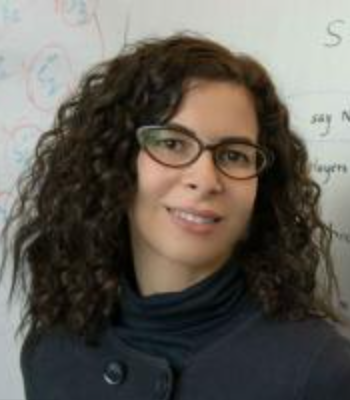}}]{Farinaz Koushanfar}
is the Siavouche Nemati-Nasser Endowed Professor of Electrical and Computer Engineering (ECE) at the University of California San Diego (UCSD), where she is the founding co-director of the UCSD Center for Machine-Intelligence, Computing \& Security (MICS). She is also a research scientist at Chainlink Labs. Her research addresses several aspects of secure and efficient computing, with a focus on robust machine learning under resource constraints, AI-based optimization, hardware and system security, intellectual property (IP) protection, as well as privacy-preserving computing. Dr. Koushanfar has received a number of awards and honors including the Presidential Early Career Award for Scientists and Engineers (PECASE) from President Obama, the ACM SIGDA Outstanding New Faculty Award, Cisco IoT Security Grand Challenge Award, MIT Technology Review TR-35, Qualcomm Innovation Awards, Intel Collaborative Awards, Young Faculty/CAREER Awards from NSF, DARPA, ONR and ARO, as well as several best paper awards. Dr. Koushanfar is a fellow of ACM, IEEE, National Academy of Inventors, and the Kavli Frontiers of the National Academy of Sciences.
\end{IEEEbiography}
\vfill

\clearpage
\newpage

\appendices

\section{Interactive vs. Non-Interactive}
\label{sec:interactive}

ZKPs can broadly be classed into two categories: interactive and non-interactive \cite{wu2014survey}. Interactive protocols, as the name suggests, require several rounds interaction before \Vrf is convinced that \Prv's proof is valid. This is done by \Vrf sending random challenges to \Prv until \Vrf is convinced that \Prv's proof is valid. Interactive ZKPs require that both \Prv and \Vrf stay online until \Vrf is convinced. This somewhat limits the utility of interactive ZKPs, as the proofs are \textit{designated-verifier}, meaning that \Prv's proof can only be used to be convinced a single verifier. A separate protocol must be performed for each new \Vrf. Conversely, non-interactive ZKPs are normally \textit{publicly verifiable}, meaning \Prv can generate a single proof in one-shot that any \Vrf can verify. Non-interactive ZKPs often rely on a trusted setup process from a third-party, or in some cases \Vrf, to generate randomness that allows for a proof to be generated that \Vrf accepts as valid without further interaction. Many non-interactive schemes aim to minimize proof size, which results in higher \Prv computational power requirements. This limits the scalability of these schemes, especially in scenarios where \Prv is resource-constrained. The interactivity of interactive ZKPs allows for a more scalable approach in terms of \Prv computation, albeit limiting the amount of verifiers that can verify a proof. If needed, there is a method for turning public-coin interactive ZKPs into non-interactive ZKPs. The Fiat-Shamir transform \cite{kilian1992note} replaces \Vrf's randomness with a random oracle (i.e. a cryptographic hash function), thus removing the interaction and turning interactive ZKPs into non-interactive ZKPs.

\section{Recursive zk-SNARKs}
\label{sec:recursive}
Recent works \cite{bowe2019recursive, kothapalli2022nova, bitansky2013recursive} have shown the usability of recursive zk-SNARKs, which is the process of verifying multiple zk-SNARKs in a single zk-SNARK. As the verification algorithm of zk-SNARKs is simply an arbitrary computation, it can be represented as a circuit \Cir. This enables one \Prv to generate many proofs, then generate a proof that verifies these proofs and send it to \Vrf. While this results in substantially more work on \Prv, \Vrf now only has to generate one proof to verify all of \Prv's data, rather than many individual proofs. 

\section{Features of ZKP Libraries}
\label{sec:app_libraries}
Table \ref{tab:description} provides a compact, high-level description of the 25 frameworks we discuss in this work.
\begin{sidewaystable*}[t!]
   \small
   \label{}
   \centering\resizebox{0.95\textwidth}{!}{
   \begin{tabular}{ccccccc}
   \toprule
   \textbf{Framework}&
   \textbf{Frontend/High-Level API} & \textbf{Dev. Language} & \textbf{Proof System(s)} & \textbf{Notable Gadgets} & \textbf{Extra Features} & \textbf{Target Audience} \\

   \midrule
    & \multicolumn{6}{c}{\textbf{zkSNARKs}} \\
   \midrule

   \textbf{Arkworks \cite{arkworks}} & Self-contained & Rust & Groth16, Marlin \cite{chiesa2020marlin}, GM17 \cite{groth2017snarky}, Plonk & Polynomial, Boolean, and UInt Arithmetic  & - &  Experienced ZK SW Developers\\
   \textbf{Gnark \cite{gnark-v0.9.0}} & Self-contained & Go & Groth16, Plonk (KZG, FRI) & Hashes, Merkle proofs, EdDSA & GPU support & SW Developers \\
   \textbf{Hyrax \cite{hyraxZK}} & None & Python & Hyrax & - & Excellent research paper &  ZK Researchers  \\
   \textbf{Zokrates \cite{eberhardt2018zokrates}} & Self-contained & Zokrates DSL & Groth16, GM17 \cite{groth2017snarky}, Marlin \cite{chiesa2020marlin}, Nova \cite{kothapalli2022nova} & Hashes, ECC & zkinterface support &  SW Developers  \\
   \textbf{LEGOSnark \cite{legosnark}} & None & C++ & Brakedown-like \cite{golovnev2023brakedown} & Sumcheck, Matrix \& vector arithmetic & - & ZK Researchers \\
   \textbf{LibSNARK \cite{libsnark}} & xjSnark \cite{kosba2018xjsnark} & Java, C++ &  Groth16, Pinocchio,  GGPR \cite{fuchsbauer2018subversion} & Hashes, Merkle trees, set commitment & Boolean \Cir support, TinyRAM & Experienced ZK SW Researchers  \\
   \textbf{Mirage \cite{Mirage}} & None & Java & Pinocchio-like & AES128, Merge Sort, SHA256 & - & ZK Researchers \\
   \textbf{PySNARK \cite{PySNARK}} & Self-contained & Python & Groth16 & Hashes, linear algebra operations  & Boolean \Cir support & Beginner ZK SW developers \\
   \textbf{SnarkJS \cite{baylina2020iden3}} & Circom \cite{munoz2022circom} & JavaScript, Circom DSL & Groth16, Plonk (via WASM) & Hashes, EdDSA, Comparators & Smart contract deployment support & SW Developers \\
   \textbf{Rapidsnark \cite{RapidSNARK}} & Circom \cite{munoz2022circom} & JavaScript, Circom DSL & Groth16 & Hashes, EdDSA, Comparators & Android/iOS \Prv support & SW Developers \\
   \textbf{Spartan \cite{Spartan}} & None & Rust & Spartan & - & Excellent research paper &  Experienced ZK SW Developers \\
   \textbf{Aurora (libiop) \cite{SciprLab2023Libiop}} & None  & C++ & Aurora & - & Excellent research paper & ZK Researchers \\
   \textbf{Fractal (libiop) \cite{SciprLab2023Libiop}} & None & C++ & Fractal & - & Excellent research paper & ZK Researchers \\
   \textbf{Virgo \cite{SunblazeUCB2023Virgo}} & None & Python & Virgo & SHA256, Lanczos algorithm & Excellent research paper &  ZK Researchers \\
   \textbf{Noir \cite{Noir2023Documentation}} & Self-Contained & Rust (Noir DSL) & Any ACIR-compatible system & Hashes, Big Integers, Merkle Trees & Recursive proof capabilities & SW Developers \\
   \textbf{Dusk-PLONK \cite{DuskPlonk2023Rust}} & None & Rust & PLONK & - & - & ZK Researchers \\
   \textbf{Halo2 \cite{Halo22023Book}} & None (Rust API) & Rust & PLONK-like & Hashes, lookup range check, field decomposition & Backend for zkML framework \textit{ezkl} \cite{ZkonduitInc2023EZKL} & Experienced ZK SW Developers \\

   \midrule
   & \multicolumn{6}{c}{\textbf{MPC-in-the-Head}} \\
   \midrule

   \textbf{Limbo \cite{KULeuvenCOSIC2023Limbo}} & EMP-tool \cite{emptool} & C++ & MPCitH & SHA256 & High level of protocol customizability & Experienced privacy SW Developers \\
   \textbf{Ligero (libiop) \cite{SciprLab2023Libiop}} & None & C++ & Ligero & - & Excellent research paper & ZK Researchers \\

   \midrule
   & \multicolumn{6}{c}{\textbf{VOLE-Based ZK}} \\
   \midrule
   \textbf{Mozzarella \cite{baum2022moz}} & None & Rust & Mozzarella & - & Excellent research paper & ZK Researchers \\
   \textbf{Diet Mac'n'Cheese \cite{dietmc}} & PicoZK & Python & Mac'n'Cheese \cite{baum2021mac} & Hashes, vector operations, histogram & Numpy, Pandas, PyTorch support & SW Developers \\
  \textbf{Emp-ZK \cite{empzk}} & Self-contained (C++ API) & C++ & Wolverine \cite{weng2021wolverine}, Quicksilver \cite{yang2021quicksilver} & Comparators, Arithmetic (e.g. log, cos) & Floating point support & SW Developers\\

   \midrule
   & \multicolumn{6}{c}{\textbf{zkSTARKs}} \\
   \midrule

   \textbf{MidenVM \cite{PolygonMiden2023MidenVM}} & None & Miden Assembly & FRI-STARK & Hashes, 64-bit arithmetic & - & ZK Researchers \\
   \textbf{Zilch \cite{TrustworthyComputing2023Zilch}} & ZeroJava \cite{mouris2021zilch} & Java & FRI-STARK & Arithmetic, logical, and bitwise operators & - & ZK SW Developers \\
   \textbf{RISC Zero \cite{RISCZero2023DeveloperDocs}} & Self-contained & Rust, C++ & FRI-STARK & Compiles any Rust code & Easy blockchain integration & SW Developers \\

   \bottomrule
   \end{tabular}}
   \caption{ZK Framework Attributes}
       \label{tab:description}
\end{sidewaystable*}



\section{ZKP Applications}
\label{sec:applications}
In this section, we discuss some of the cutting-edge ZKP applications that have been introduced in academia and industry. For an extensive view on the more simplistic applications of ZKPs, we refer readers to the excellent work of \cite{Chainlink2023ZeroKnowledgeProof}. 

\textbf{Verifiable Machine Learning. }
Verifiable computation (VC) is a technique enabled by ZKPs that allows one party to prove to another that computation was performed correctly and soundly without revealing any information about the underlying data or computation details \cite{vsimunic2021verifiable}. This is most common when there is a computationally weak verifier that would like to outsource their computation to a strong prover. This scenario lends itself quite nicely to verifiable machine learning (VML), in which a verifier can outsource their inference to a prover who owns a proprietary model. Many academic \cite{weng2021mystique, liu2021zkcnn, lee2020vcnn, feng2021zen} and industry \cite{ZkonduitInc2023EZKL} works have enabled VML, in which a cloud server (the prover) provides a ZKP that attests to the verifier that inference was computed soundly, without revealing any information about the server's proprietary model.


\textbf{zk-Rollups. }
One of the biggest problems that faces the widespread implementation of ZKPs in modern systems is the difficulty of scalability. This is evident in blockchain applications, like Zcash \cite{hopwood2016zcash} and Monero \cite{Monero2023}, which require heavy computational efforts to protect each transaction on the blockchain that they hope to keep private. zk-Rollups aim to address similar problems, although not specific to Zcash and Monero, by aggregating multiple transactions into a single batch and generating a single proof that validates all of them in one shot. This is mostly enabled by the use of recursive zk-SNARKs \cite{kothapalli2022nova}, in which a ZKP for each transaction is built, followed by a ZKP that validates all of the transactions at once. This significantly lightens the computational load on the verifier. zk-Rollups have become a more prominent solution towards applying ZKPs at scale on the blockchain in several industrial efforts \cite{0xPolygonZero2023Plonky2, PolygonLabs2023PolygonZkEVM}.

\textbf{Robust Federated Learning. }
Byzantine attacks on federated learning refer to a security threat in which malicious users aim to harm the central model \cite{fang2020local}. The introduction of secure aggregation \cite{bonawitz2017practical}, which was devised to secure individual user updates, has made it much easier for malicious users to perform Byzantine attacks. In secure aggregation, malicious users can simply hide amongst benign users and inject poisoned updates that affect the central model's accuracy. Even if a malicious attack is detected, the privacy-preserving nature of secure aggregation, the attacker cannot be identified. Several academic works \cite{ghodsi2023zprobe, so2020byzantine, roy2022eiffel, lycklama2023rofl} have proposed scalable and secure secure aggregation schemes that utilize ZKPs to check individual user gradients, allowing for detection and exclusion of malicious users, while still maintaining end-to-end privacy.

\textbf{Digital Signatures. }
In the search for more post-quantum secure digital signatures, the National Institute of Standards and Technology (NIST) began a search for additional schemes to standardize in 2023 \cite{NIST_PQC_Round2_Signatures}, following up their previous digital signature standardization efforts. Their goal was to identify lightweight digital signature schemes that maintain high privacy and security in the presence of quantum adversaries. In particular, NIST called for general-purpose schemes that do not rely on lattices and maintain fast verification and short signature size. ZKPs have proven to be an excellent cryptographic primitive for the digital signature schemes that have found success in the NIST standardization process. Recently, in October 2024, NIST announced 14 second-round candidates for post-quantum digital signatures. Out of the 14 candidates, 6 of the candidates are built using ZKP schemes. The schemes Mirath \cite{PQC_MIRATH_Website}, MQOM \cite{benadjila2024mq}, PERK \cite{bettaieb2024perk}, RYDE \cite{aragon2023ryde}, and SDitH \cite{SDITH_Website} all utilize the MPCitH ZK scheme, while FAEST \cite{baum2023publicly} uses VOLE-in-the-head, a non-interactive implementation of VOLE-based ZK. These works are currently undergoing a thorough cryptanalysis and evaluation process before they are advanced to standardization. These efforts highlight the effectiveness of ZKP schemes, especially non-interactive ones, in real-world and large-scale applications. ZKPs serve as the perfect underlying technology for post-quantum digital signature due to their public verifiability and succinct proof sizes, which enables fast verification at scale.

\textbf{FHE Integrity. }
Similar to VML, FHE integrity consists of a verifier sending their data to a cloud server for computation. However, in FHE, the computation is done on encrypted data, making the ZKP generation process much more complex. As FHE operations are more computationally intensive and use underlying ring arithmetic, the circuit that expresses the computation for a ZKP grows to be very complex. \cite{ganesh2023rinocchio} introduces a ring-based zk-SNARK enabling verifiable computation over encrypted data, however new works \cite{viand2023verifiable} have shown that, although this make FHE integrity proofs feasible, the overhead makes it an impractical solution. 

\textbf{Data Authenticity. }
ZKPs have been integrated into DECO \cite{zhang2020deco}, a protocol that is being used in practice by Chainlink Labs \cite{breidenbach2021chainlink} which allows users to prove the authenticity of their data without revealing any information about the data itself, including the length of the datapoints. DECO allows users to prove that their data was sourced from a legitimate location, while also allowing them to prove certain attributes of the data (e.g. proving account balance is above a certain threshold). Proofs surrounding data authenticity, validity, and attributes lend themselves very nicely to ZKP settings. Several domains, such as healthcare and finance, which Chainlink has shown feasibility of, can benefit from integrating ZKPs to protect user data while still gathering meaningful information. Recent work \cite{juels2024props} has shown the value of using ZKPs to build protected pipelines, or \textit{props} for short, to provide verifiable, privacy-preserving access to deep web data for machine learning pipelines. This kind of secure access to deep web data is an integral part of advancing machine learning paradigms, as it enables developers to bypass the bottleneck of limited high-quality training data that is not currently accessible.

\end{document}